\documentclass[preprint]{aastex}

\usepackage{color}

\newcommand{\nix}{$\cdot\cdot\cdot$}
\newcommand{\Rg}{\ensuremath{\mathrm{r_{g}}}}
\newcommand{\Msun}{\ensuremath{\mathrm{M_{\odot}}}}  % produces upright M
\newcommand{\NH}{\ensuremath{\mathrm{N_{H}}}}        % upright N and H in Nh
\newcommand{\cgsflux}{\ensuremath{\mathrm{ergs \ cm^{-2} \ s^{-1}}}}
\newcommand{\mathsub}[2]{\ensuremath{\mathrm{#1_{#2}}}}
\slugcomment{To appear in the Astrophysical Journal.}

\shortauthors{Weaver, Gelbord \& Yaqoob}
\shorttitle{Variable Fe lines in Seyfert~1s}

\begin{document}

\title{Variable Iron K$\alpha$ Lines in Seyfert~1 Galaxies}

\author{K. A. Weaver\altaffilmark{1,2}, J. Gelbord\altaffilmark{2}
       and T. Yaqoob\altaffilmark{1,3}}

\altaffiltext{1}{Laboratory for High Energy Astrophysics,
NASA/Goddard Space Flight Center, 
Greenbelt, MD 20771}
\altaffiltext{2}{Johns Hopkins University, Department of Physics and
Astronomy, Homewood campus, 3400 North Charles Street,
Baltimore, MD 21218}
\altaffiltext{3}{Joint Center for Astrophysics, Physics Department,
University of Maryland, Baltimore County, 
1000 Hilltop Circle, Baltimore MD 21250.}

\begin{abstract}

We find that variability of the iron K$\alpha$ line is common in
Seyfert~1 galaxies.  Using data from the {\it ASCA} archive for objects
that have been observed more than once during the mission, we study the
time-averaged spectra from individual observations, thereby probing
variability on timescales that range from days to years.  Since the
statistics of the data do not warrant searches for line variability in
terms of a complex physical model, we use a simple Gaussian to model the
gross shape of the line, and then use the centroid energy, intensity and
equivalent width as robust indicators of changes in the line profile.
We find that $\sim$$70\%$ of Seyfert~1s (ten out of fifteen) show
variability in at least one of these parameters: the centroid energy,
intensity, and equivalent width vary in six, four, and eight sources
respectively.  Due to the low S/N, limited sampling and time averaging,
we consider these results to represent lower limits to the rate of
incidence of variability.  In most cases changes in the line do not
appear to track changes in the continuum.  In particular, we find no
evidence for variability of the line intensity in NGC~4151, suggesting
an origin in a region larger than the putative accretion disk, where
most of the iron line has been thought to originate.  Mkn~279 is
investigated on short timescales.  The time-averaged effective line
energy (as measured by the Gaussian center energy, which is weighted by
emission in the entire line profile) is 6.5~keV in the galaxy rest
frame.  As the continuum flux increases by $20\%$ in a few hours, the
Fe~K line responds within $\sim$10,000~seconds, with the effective line
energy increasing by 0.22~keV ($\sim$10,500~km~s$^{-1}$).  We also
examine the {\it Rosat} PSPC spectrum of Mkn~279 but find
inconsistencies with {\it ASCA}.   Problems with the {\it ASCA} and
{\it Rosat} calibration that affect simultaneous spectral fits at low
energies are discussed in an appendix.

\end{abstract}


\section{Introduction}

Active galactic nuclei (AGN) are thought to be powered by the release
of gravitational energy as matter falls onto a supermassive black
hole.  There is currently much evidence supporting this scenario
(e.g., see Collin et~al.\ 2000, Mushotzky, Done \& Pounds 1993), but  
the source of the accreting material and how it makes its way into the
center of the galaxy remains uncertain.  One important piece of the
puzzle is understanding how this material is
distributed.  The current AGN paradigm invokes an
accretion disk that extends from approximately the innermost stable
orbit of the black hole out to a few hundredths pc in radius
($10^4 \; \Rg$ for a $10^8 \; \Msun$ black hole, where $\Rg = GM/c^2$
is the gravitational radius.).  The inner disk is believed to be
surrounded by a much larger structure, a few parsecs across.  This
structure feeds the inner disk, and serves as the obscuring torus in
unification schemes (Antonucci 1993),  in which the main difference
between Type~2 (narrow-line) and Type~1 (broad-line) objects is due to
their orientation with respect to the observer (e.g., Antonucci \&
Miller 1985; Madejski 1999).  The current most advanced telescopes
cannot image gas disks smaller than a few hundred pc across and so
studying AGN on sub-parsec scales requires spectral techniques.  One
such technique involves measuring the relationship between emission-line
and continuum variability.  This technique, referred to as reverberation
mapping, allows us to infer the sizes and kinematics of regions that
reprocess continuum radiation.

To derive the physical properties of the accretion disk and/or other
matter that is in the vicinity of the black hole, X-ray observations
are crucial.  The proximity of the inner edge of the disk to the black
hole exposes it to extreme conditions of temperature, ionization and
gravity.  This causes it to be X-ray bright, not only in emission but
also via reprocessing continuum photons.  The most robust feature of
X-ray reprocessing is the Fe~K$\alpha$ fluorescence line.  The light
crossing time at the innermost stable orbit around a Schwarzschild black
hole is $\sim$3000~M$_8$~s (where M$_8 = \mathrm{M_{BH}} / 10^8 \; \Msun$).
Hence, an Fe~K$\alpha$ line produced by the inner disk should respond
rapidly to changes in the continuum emission, on timescales as short as
a few hours.  Emission at 100~\Rg, on the other hand, can have response
times as large as 50,000~s.  By studying changes in the line intensity,
width and skewness, as well as searching for time lags/leads in the
Compton reflection component, we can infer important physical properties
of the accreting matter, the central black hole and the size and
geometry of the Comptonizing medium responsible for the hard X-ray power
law (e.g., Reynolds 2000).

So far, time-resolved spectroscopy of the Fe~K$\alpha$ line in AGN is
scarce.  Of past X-ray missions, only the {\it ASCA} satellite has had
sufficient energy resolution and observing efficiency at $\sim$6~keV
to begin to resolve the iron line profile.  Many AGN have been observed
with {\it ASCA}, but the large ratio of signal to noise that is needed
for reliable time-resolved spectroscopy of the line is attained for only
a few objects.  There are only four Seyfert~1 galaxies for which the
data have allowed investigation of short-term variability of the shape
of the Fe~K$\alpha$ line profile: MCG~--6-30-15 (Iwasawa et~al.\ 1996),
NGC~7314 (Yaqoob et~al.\ 1996), NGC~3516 (Nandra et~al.\ 1999), and
NGC~4051 (Wang et~al.\ 1999b).  All show variability on timescales less
than 30~ks, but the behavior of the line is complex, with some portions
responding to continuum variations and others not.  This complex pattern
is not easy to understand within the framework of pure reverberation
(Reynolds 2000).  Complicating things further, the line from the disk
may be convolved with emission from further out, making deblending
important (Weaver \& Reynolds 1998).

The best AGNs in which to study line variability are those for which
there is reason to believe that the Fe~K$\alpha$ line is dominated by
emission from the accretion disk.  This is most likely the case for
Seyfert~1 galaxies, where the mean line profile shows significant
Doppler and gravitational broadening, suggesting a strong contribution
from the accretion disk (Nandra et~al.\ 1997; hereafter N97).  In this
paper, we use {\it ASCA} observations of Seyfert~1 galaxies to examine
the variability properties of the iron line.  Our goal is to determine
whether variability is common and what it may tell us about the
structure of the central engine.

We begin our discussion with a detailed look at Mrk~279 (\S2).  In this
galaxy, we found that the continuum flux changed significantly halfway
through the observation and we were prompted to investigate how the Fe-K
line responded to this continuum change. Given the limited statistical
quality of the data we were motivated to use the emission-weighted
centroid (as modeled by a Gaussian) of the line as one of the primary
indicators of variability of the line profile.  Realizing that this is a 
powerful probe of line shape variability for limited signal-to-noise
data, we launched a program to look for Fe-K line variability across a
{\em sample} of Seyfert~1 galaxies, rather than focusing on just the few
brightest sources.  Our sample is defined as all Seyfert~1s with
multiple {\it ASCA} observations that have been archived in the
HEASARC\footnote{The High Energy Astrophysics Science Archive Research
Center at NASA/Goddard Space Flight Center.} database as of 1999 August. 
In \S3, we apply this new model to the sample of fifteen Seyferts, which
does not include Mrk~279 (because it has only been observed once).  For
this pilot study, we only consider the time-average spectrum of each
observation.  A discussion of results follows in \S4, and our results
are summarized in \S5. In the process of investigating the broad-band
X-ray spectrum of Mrk~279, we encountered significant discrepancies
between the {\it ASCA} SIS (solid state spectrometer), GIS (gas imaging
spectrometer) and the {\it Rosat} PSPC (position sensitive proportional
counter).  These are discussed in an appendix along with a prescription
for correcting some time-dependent effects in the SIS data.

\section{Mrk~279}

\subsection{Observations}

Mkn~279 was observed with {\it ASCA} on 5 December 1994 with the SIS
in 2-CCD mode.  Using the most recent calibration, preliminary analysis
confirmed the known problem of a discrepancy between the lowest-energy
response of SIS1 compared with the other three instruments, so we
excluded the SIS1 data below 1~keV from our analysis.  The remaining SIS
data are obtained in bright mode and times of high background are
removed based on the criteria recommended in the {\it ASCA} data
analysis guide\footnote{Hot and flickering pixels are discarded, as are
data when the satellite passes through the SAA, when its elevation above
Earth's limb is $<5^{\circ}$ (night) or $<20^{\circ}$ (day), when the
geomagnetic cutoff rigidity is $<6$~GeV~c$^{-1}$, and when the time is
$<32$~s after a day/night transition, or $<32$~s after an SAA passage.}.
We include grade 6 photon events in the SIS to improve the ratio of
signal to noise at high energies (see e.g., Weaver et~al.\ 1996).
Approximately 22~ks of good data are obtained from each SIS and 25~ks
from each GIS, resulting in 2--10~keV count rates of 0.95, 0.81, 0.56,
and 0.69~counts~s$^{-1}$ in S0, S1, G2 and G3, respectively.

Mkn~279 was observed with the {\it Rosat} PSPC on 1993 October 7 for
approximately 10~ks.  The processed data are obtained from the HEASARC
and the source spectrum is extracted with the {\it ftools} software
package.  Background is accumulated from source-free regions near the
target and then subtracted.  The {\it Rosat} and {\it ASCA} spectra
are grouped to have $\ge30$ photons in each energy bin and are modeled
separately with the {\it xspec} spectral fitting package.

\subsection{Temporal Variability}

\begin{figure}
\epsscale{0.6}
\plotone{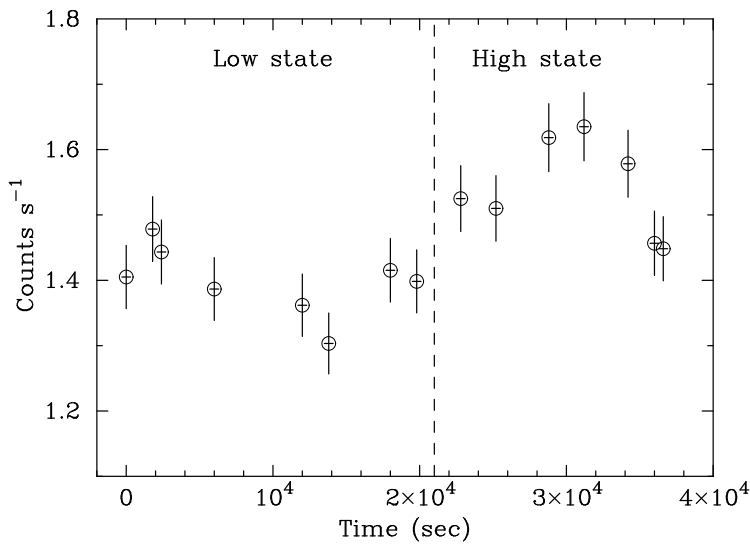}
\caption{{\it ASCA} SIS0 light curve of Mkn~279.  The data are binned 
into 600~s intervals.  The times marked ``Low'' and ``High''
correspond to the intervals selected for time-resolved spectra.
\label{fig:279s0LC}}
\end{figure}

Mrk~279 exhibits X-ray variability on timescales ranging from hours to
years.  Rapid variability is evinced by the {\it ASCA} observation,
during which the source varied in amplitude by $\sim$$20\%$
in $\sim$10~ks.  Modeling the S0 lightcurve in Figure~\ref{fig:279s0LC}
with a constant yields $\chi^2/\nu = 3.36$, which rules out a constant
source at $\gg 99\%$ confidence.
 
\begin{figure}
\plotone{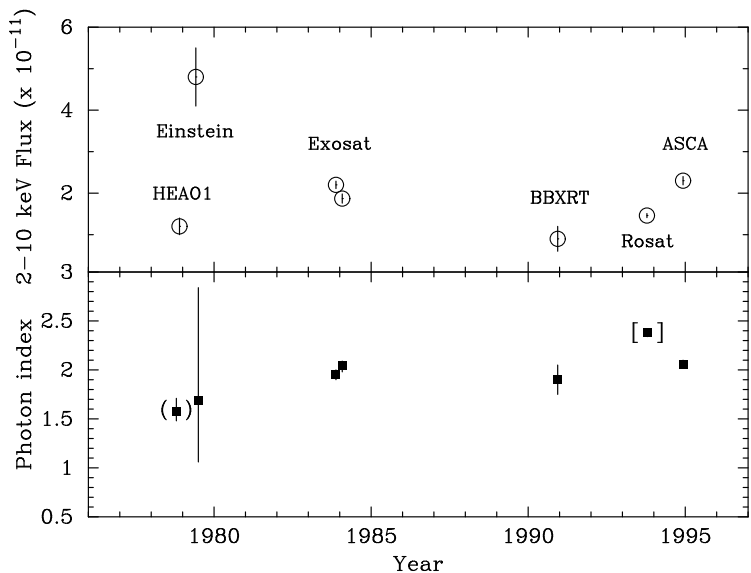}
\caption{The 2--10~keV flux and photon index derived from all pointed
X-ray observations of Mkn~279 from 1979 through 1994.  The photon
index is determined from $\sim$2--10~keV, except for the {\it Rosat}
value (in brackets), which is measured from 0.2--2.4~keV.  The 
{\it Rosat} flux is estimated by extrapolating the power law to 10 keV.  
The {\it HEAO-1} measurement (in parenthesis) is weighted toward slightly
higher energies and does not account for any reflection component.
\label{fig:279FAndGammaVar}
}
\end{figure}

Pointed X-ray observations of Mrk~279 from 1979 to 1994 indicate
long-term source variability of up to a factor of~5
(Figure~\ref{fig:279FAndGammaVar}).  Throughout this entire period the
derived photon index remains fairly stable at a value of $\Gamma = 1.9$.
The exceptions are measurements from {\it HEAO-1} A2 (parenthesis) and
{\it Rosat} (brackets).  The larger {\it Rosat} index may result from a
soft ``excess'' emission component that becomes important below
$\sim$0.6~keV (\S2.3), while the smaller {\it HEAO-1} A2 index may be
caused by Compton reflection that flattens the spectrum above
$\sim$7~keV.  Upon re-examination, the {\it HEAO-1} data published by
Weaver, Arnaud and Mushotzky (1995) are consistent with having an
underlying photon index of 2.0 when Compton reflection from a disk with
a covering factor of $\Omega = 2\pi$ is included in the ``A2'' spectral
model.
                                  
\subsection{The Time-Averaged {\it ASCA} and {\it Rosat} PSPC Spectra}

\begin{figure}
\plotone{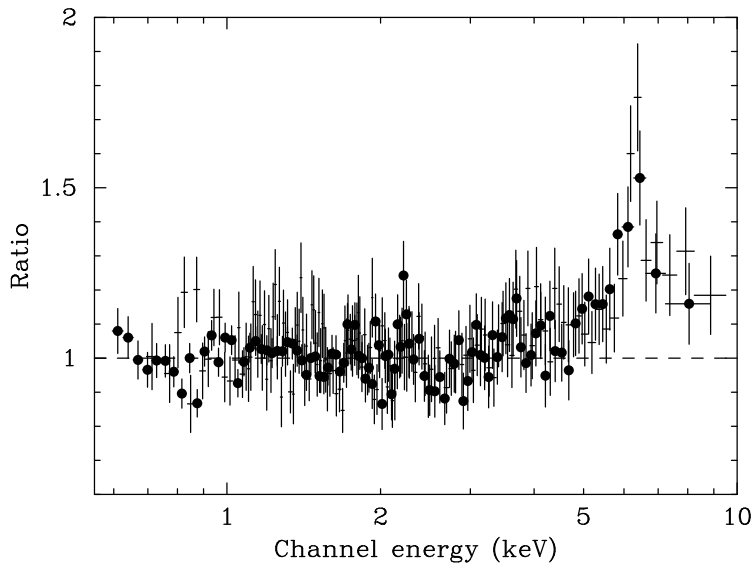}
\caption{Ratio of the {\it ASCA} data to a model that consists of a
power law spectrum absorbed by the Galaxy.  The Fe~K$\alpha$ emission
line is clearly present as well as the suggestion of an upturn
(flattening) in the continuum above $\sim$6~keV.  The upturn is
consistent with the expected amount of Compton reflection from an
accretion disk producing the line.
\label{fig:279modelRatio}
}
\end{figure}

We first examine the time-averaged spectrum and discuss in turn the
following three energy bands: 0.6--1.0~keV, 1--4~keV, and 0.6--10~keV.
At energies of 0.6--1.0 and 1--4~keV, Mkn~279 is bright and featureless,
and so we can place stringent limits on the shape of the underlying
continuum.  The {\it Rosat} data show no evidence for absorption
exceeding that from our Galaxy and so we fix the absorption at the
Galactic value of $\NH\ = 1.6\pm0.3\times10^{20}$~cm$^{-2}$ for all
spectral fits (Elvis, Lockman \& Wilkes 1989).  Fitting the data from
0.6 to 1~keV with a model that consists of an absorbed power law yields
a photon index of $2.1\pm0.1$ ($90\%$ confidence error\footnote{Unless
otherwise noted, statistical errors reported for all measurements are 90\%
confidence for one interesting parameter (corresponding to
$\Delta\chi^2 = 2.706$).}).  Going to higher energies and fitting the
same model from 1~to 4~keV yields a similar index of $\Gamma =
2.06\pm0.02$.  The equality of spectral slope across the entire 0.6 to
4~keV band indicates no evidence for a spectral upturn below 1~keV in
the {\it ASCA} spectrum (what otherwise might be called a soft
``excess'').  Considering the entire energy band, a power law provides a
statistically good fit (Table~\ref{tab:integrated279}, model~1;
$\chi^2/\nu = 923/860$), but significant deviations from the power law
shape are evident above $\sim$5~keV (Figure~\ref{fig:279modelRatio}).

The iron K$\alpha$ line is clearly detected.  Adding a narrow Gaussian
to model the line (model~2) yields $\Delta\chi^2 = 57.5$ compared to
model~1, which represents a highly significant improvement at
$\gg$$99\%$ confidence for the addition of two free parameters.  Adding
another free parameter by allowing the width of the line to vary
(model~3) increases the uncertainty in the line energy, but improves the
fit further with $\Delta\chi^2 = 18.8$, again significant at $\gg$$99\%$
confidence.  Replacing the Gaussian model with an emission line from a
disk around a non-rotating black hole (Schwarzschild metric--- Fabian
et~al.\ 1989) provides a slightly better overall fit (model~4), but in
this case requires the disk to be nearly edge-on.  However, we found
that this broad, large disk line is actually modeling the high-energy
excess in the continuum rather than the sharp, narrow line feature (see
Figure~\ref{fig:279modelRatio}); such an inclination is unlikely given
the unabsorbed nature of the continuum source, unless there is a
significant misalignment of the accreting (inner) disk and outer disk or
torus.  A more compelling reason against an edge-on disk is that it
would have great difficulty producing the required, large equivalent
width (George \& Fabian 1991).
 
\begin{figure}
\plotone{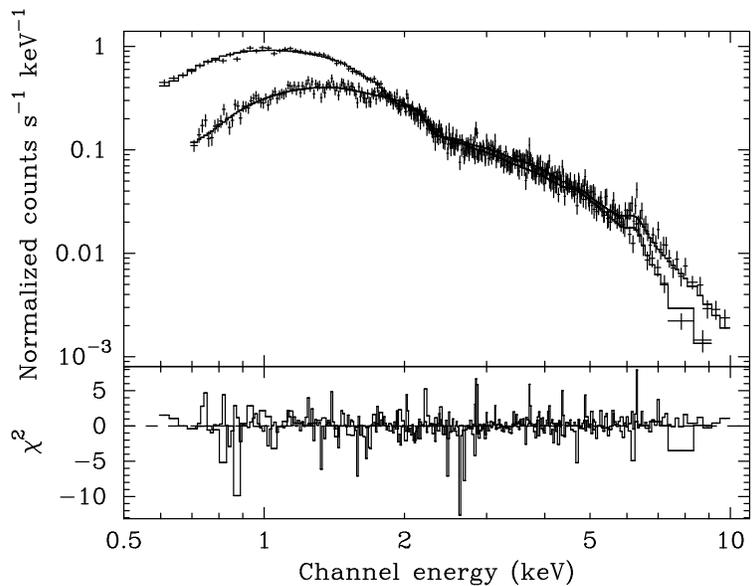}
\caption{Top panel: The {\it ASCA} data and spectral model folded
through the instrumental responses.  Only data from S0 and G3 are shown
for clarity. The model contains an Fe~K line and Compton reflection
(Table~1, model~9).  Bottom panel: $\chi^2$.
\label{fig:279spectrum}
}
\end{figure}

We next examine models that include Compton reflection (Magdziarz \&
Zdziarski 1995).  This choice is physically motivated but has its
limitations, because {\it ASCA} cannot constrain the reflection spectrum
well due to its limited bandpass and high-energy effective area, so the
amount of reflection is consequently model-dependent.  A model with
reflection from neutral material and no iron line (model~5) provides a
fit that is comparable to models with a broad Gaussian or a disk line
($\chi^2/\nu = 849/859$).  However, the best fits include both Compton
reflection {\it and} an iron line (models 6, 7, 8 and 9).  Comparing
$\chi^2$, we see that the data cannot distinguish unambiguously between
a narrow and broad line because there is a severe degeneracy between the
line width and the reflection normalization $R$ ($\chi^2/\nu = 825.8/857$
for a narrow Gaussian and 824.3/856 with $\sigma = 0.14\pm.14$~keV when
the width is a free parameter).  If a canonical amount of reflection is
assumed ($R = 1$ for a disk with $\Omega = 2\pi$) then the line is
barely resolved with $\sigma \simeq 0.35 \pm 0.25$~keV (model~8).
Including reflection and then fitting the Fe~K$\alpha$ line with an
emission feature from a disk around a non-rotating black hole yields a
disk inclination of $\sim$$30^{\circ}$ (model~9;
Figure~\ref{fig:279spectrum}), as opposed to the large inclination found
previously when not including reflection (model~4).  The smaller
inclination is similar to the mean value for Seyfert~1s (N97),
suggesting that the reflection model is a reasonable assumption.  Note,
however, that this model is statistically indistinguishable from
Gaussian models~6 and~7.

The {\it Rosat} PSPC and {\it ASCA} spectra are not consistent.  The
agreement between column densities is excellent for the PSPC and
SIS0\footnote{But see appendix for a comparison including SIS1.}, but
the PSPC data require a power law having $\Gamma = 2.38$.  This index
is larger by a factor of $\Delta\Gamma \sim 0.4$ compared to the
{\it ASCA} index of $\Gamma = 2.0$.  The difference in the photon
indices may result from an {\it ASCA} miscalibration, it may be caused
by a remaining PSPC calibration error (Iwasawa, Fabian and Nandra 1999;
Snowden, Turner \& Freyberg 2000) or there may be a separate component
that dominates the PSPC spectrum.  Apparent steepening below
$\sim$0.6~keV could also result from a variable soft X-ray component,
since the {\it ASCA} and PSPC observations are not simultaneous.
Because we cannot determine the true source of the discrepancy, we do
not discuss joint PSPC and {\it ASCA} fits.  The {\it ASCA} and
{\it Rosat} cross-calibration is discussed further in the appendix.

\subsection{The High- and Low-State {\it ASCA} Spectra}

Dividing the {\it ASCA} data for Mkn~279 according to the high and low
flux states indicated in Figure~\ref{fig:279s0LC} allows a search for
spectral changes on $\sim$10,000~s timescales.  There is no evidence for
a change in photon index or amount of reflection between the two
time-resolved spectra, although we cannot rule out changes in the
Compton hump as large as $20\%$.  On the other hand, the Fe~K line
appears to vary, although modeling the entire, broadened profile is
difficult and the line width and normalization depend on how the
continuum is modeled.

In order to make quantitative statements about the variability of the
iron line we have taken steps to determine which {\it ASCA} measurements
are robust.  Because of the limited energy resolution, bandpass and S/N
of these data, which make it hard to know the shape of the underlying
continuum, it is difficult to quantify the shape of broad spectral
features.  Such is not the case for narrow features, even between models
that include or exclude Compton reflection.  Therefore we have chosen a
method of using a simple {\it empirical} model of a single, unresolved
Gaussian ($\sigma = 0.05$~keV) to model the Fe~K line, with the
assumption that the energy of this Gaussian reflects the weighted peak
of the line profile.  The strength of this method is that it provides us
with a single, model-independent parameter that we can trust.  Any
change in the effective line energy indicates that the line profile is
changing.  Furthermore, this method enables us to significantly increase
the number of sources for which we have some iron line variability
information from a few to a sizable sample (\S3).

The weakness of this method lies in the fact that we are recording a
weighted average of the line profile, not necessarily a real change in
the peak energy.  Therefore we cannot know {\it how} the line changes,
i.e., whether the energy changes or whether redshifted emission
vanishes leaving a blueshifted peak or whether there are absorption or
other effects.  However, when an energy shift is measured, we can say
with confidence that the line profile {\it has} changed.

\begin{figure}
\plotone{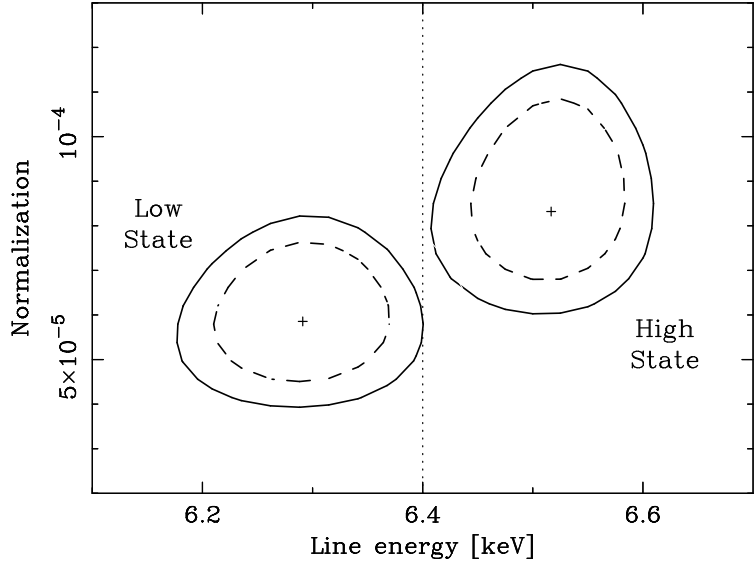}
\caption{Gaussian line energy vs. normalization for the Fe~K$\alpha$
line in the low (left) and high (right) flux states.  The dashed and
solid contours respectively represent the $68\%$ and $90\%$ confidence
levels (for two interesting parameters).  The vertical line represents
emission from neutral material with no Doppler effects and is drawn as a
reference.
\label{fig:mergedCont}
}
\end{figure}

For Mkn~279 using a single Gaussian, we find that the effective peak
energy of the Fe~K$\alpha$ line changes in $\sim$10,000~s by
$\sim$$3\%$, increasing from 6.30$^{+0.10}_{-0.12}$~keV in the 
low state to 6.52$^{+0.09}_{-0.09}$~keV in
the high state (Figure~\ref{fig:mergedCont}).  If interpreted as a
simple Doppler shift, this corresponds to a velocity shift of
$\sim$10,500~km~s$^{-1}$.  We conclude that there is a significant
change in the line profile in $\sim$10,000~s.

\section{The {\it ASCA} Seyfert~1 Sample}

Having established a simple, reliable way to characterize basic
variability of the Fe~K line profile, we turn to the {\it ASCA}
sample.  This sample is composed of all {\it ASCA} observations of all
Seyfert~1s with multiple pointings available in the HEASARC public
archives as of August 1999, and includes 72 pointings towards 15
galaxies.  In an effort to survey this sample for line variability, we 
apply the approach described for Mkn~279.  Unlike our study in \S2.4, we 
limit this initial {\em pilot} program to comparisons of the time
average spectra of the observations, looking only for variations from
one pointing to another.  This provides a reliable measure of
variability on time scales ranging from days to years, but does not
address more rapid changes, such as the $\sim$10~ks variability found in
Mkn~279.  We leave the study of line variability on intra-observation
timescales to future work.

\subsection{Spectral Fits}

\begin{figure}
\figurenum{6}
\epsscale{1.1}
\plottwo{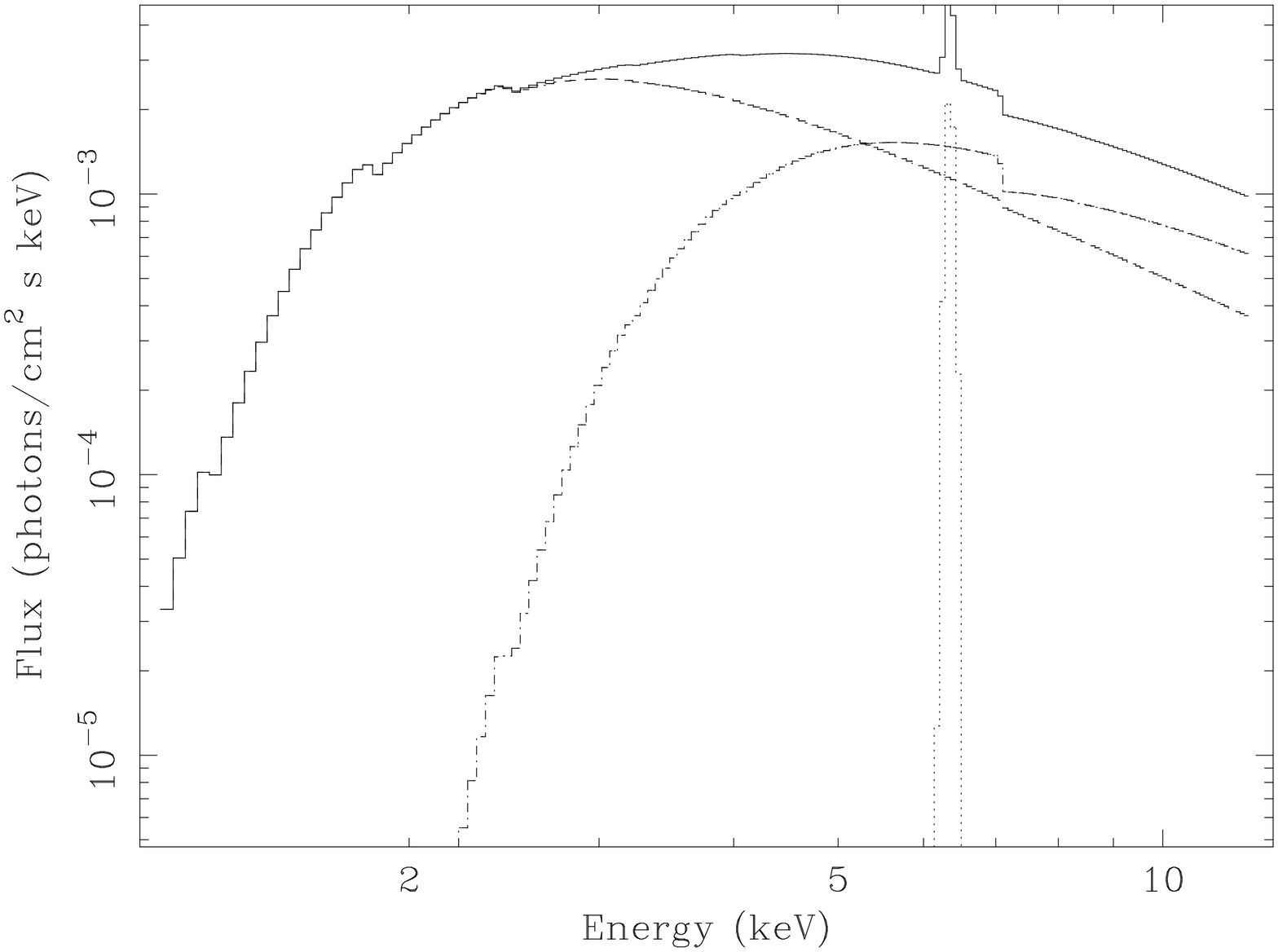}{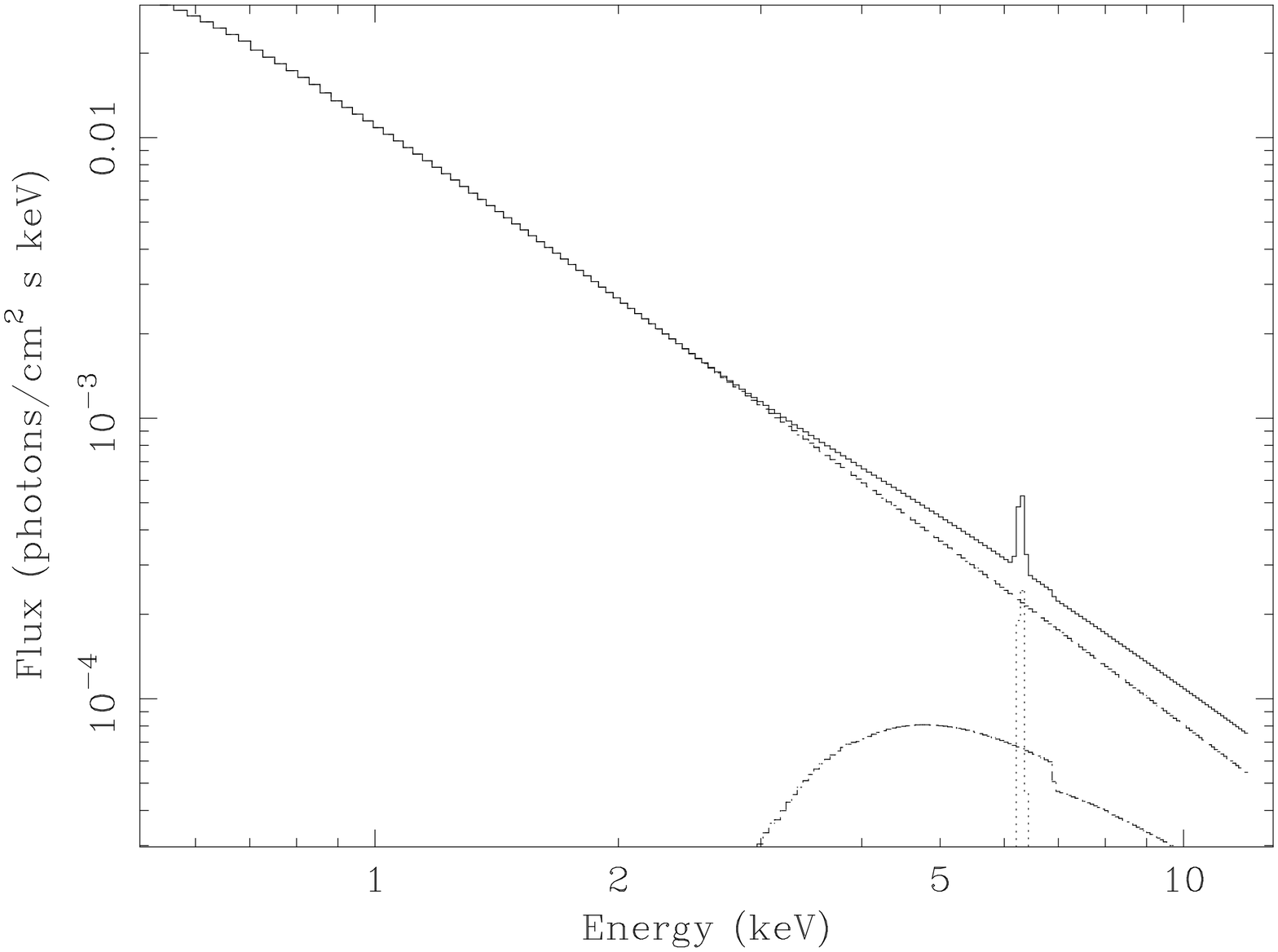}
\caption{Our general empirical model, as applied to the spectra of two
extreme galaxies.  To the left are the best-fit model components for
NGC~4151, and to the right are the best-fit components for Mkn~279.  The
strong, broad continuum feature above $\sim$4~keV in NGC~4151 appears to
be the intrinsic source which is partially absorbed by material with a
column density of $\sim5\times10^{22}$ cm$^{-2}$, while the weaker broad
feature above $\sim$4~keV in Mkn~279 is most likely produced by Compton
reflection.
\label{fig:model_applied}
}
\end{figure}

The data are fitted from 1 to 10~keV, using a spectral model consisting
of two power laws and a narrow Gaussian.  In the galaxy rest frame the
spectrum is given by: \\
\begin{displaymath}
I = N_1 E^{-\Gamma} 
	+ N_2 e^{- \mathrm{N_H \sigma_{abs}}(E)} E^{-\Gamma}
	+ N_3 (2\pi\sigma^2)^{-1/2} e^{-0.5[(E-E_o)/\sigma]^2}
	\ \mathrm{photons\;s^{-1}\;cm^{-2}\;keV^{-1}}
\end{displaymath}
where $\sigma_{\mathrm{abs}}(E)$ is the photoelectric cross-section and
$N_1$, $N_2$, and $N_3$ are independent normalizations for each
component.  When this model is fit to the data, an extra absorbing
column at $z=0$ is applied to the entire spectrum.  This column density
is a free parameter, but is not allowed to be less than the Galactic
\NH\ value.  We fix the Gaussian width at $\sigma = 50$~eV, which is
narrower than the best {\it ASCA} spectral resolution in the Fe~K
region.  This representation of the core of the emission line is simply
used to characterize the peak energy of the line emission and the flux
in the unresolved core of the line.  For most galaxies, broadening of
the iron line is at best only marginally detectable because of the low
signal-to-noise ratio of the data; a narrow Gaussian only grossly
underrepresents the {\it ASCA} profile for a few cases, such as
MCG~--6-30-15 (see N97), NGC~4151 (see Yaqoob et~al.\ 1995, Wang, Zhou
\& Wang 1999a) and NGC~3516  (see Nandra et~al.\ 1999).  The continuum
model represents an intrinsic underlying power law which intercepts an
absorber with a covering fraction equivalent to the ratio of the
normalization of the obscured power law to the total continuum
normalization.  Physically this may either correspond to the source
being observed in transmission through a partially covering absorber or
the source being observed in transmission through a fully covering
absorber but with some fraction of the direct continuum being observed
by optically thin reflection out of the line of sight.  The signature of
such a physical scenario upon a spectrum is illustrated in
Figure~\ref{fig:model_applied} (left panel).  However, a high-energy
excess could also be produced by Compton-thick reflection, whereby
continuum emission is back-scattered off a Compton thick surface.  The
signature of Compton-thick reflection is strongest at higher energies
(20--50~keV); in the {\it ASCA} bandpass it is often difficult to
constrain, but it can be parameterized by our second, heavily absorbed
power law component.  Our model is meant to be a ``universal'' model
that allows us to treat all of the galaxies in our sample identically;
it does not require assumptions about the physical nature of the broad,
high-energy continuum features.  Complex photoelectric absorption is the
likely origin of the hard X-ray bump in some targets, such as NGC~4151
(Weaver et~al.\ 1994), while Compton-thick reflection creates the excess
in several others, including Mkn~279 (\S2).  The application of our
model to these two extreme cases is illustrated in
Figure~\ref{fig:model_applied}.

Note that the EW values reported here are not necessarily comparable to
those reported elsewhere in the literature.  They are based upon the
{\em observed} flux of the line core and the total X-ray continuum,
which includes the {\em absorption corrected} second power law.  The
absorption column on the second power law is generally not very well
constrained, and can occasionally be quite large.  In such cases, the
absorption correction can be substantial and hence the EW values can be
small.  Thus, we consider our relative EW values to be more meaningful
than absolute values and therefore focus our discussion upon relative
comparisons for the same source.  The spectral fits are presented in
Table~\ref{tab:sample}.

Spectral results for Seyfert~1 galaxies have been published by Nandra
et~al.\ (N97), who presented time-averaged iron line profiles from
observations obtained during the first two years of the {\it ASCA}
mission\footnote{N97 uses 23 observations of 18 Seyferts.  Four of these
targets were not observed again, so they are not part of our sample.}.
To make sure our modeling technique has not introduced spurious effects
we compare our results with those of N97, who model the 3--10~keV
spectrum with a power law plus a narrow Gaussian and no hard X-ray
absorption or reflection.  In Figure~\ref{fig:lineEdist}, we compare the
distributions of line center energies from N97 with our measurements.
Our results are presented in the histogram, with shaded bins denoting
observations used by N97.  The curves are the best fit Gaussian
distributions for the N97 results (dotted curve; $6.34\pm0.12$~keV),
compared with our results for the N97 sample (dashed curve;
$6.37\pm0.09$~keV), and for pointings in either sample (solid curve;
$6.38\pm0.11$~keV; only 64 observations are represented because there
was no significant iron line in the other 12).  Our measurements compare
favorably with the N97 results, both in the mean and individually.

\subsection{The Fe~K$\alpha$ Line}

The emission line and continuum parameters are plotted in
Figure~8a--o for all galaxies in our
sample.  For Fairall~9, NGC~4151, MCG~--6-30-15 and NGC~5548, we also
show correlation plots of interest in Figure~9a--d.

Of the fifteen galaxies with more than one {\it ASCA} pointing,
fourteen show significant 2--10~keV variability\footnote{We say that a
parameter has varied between two observations when the statistical
errors for those measurements do not overlap.  Our confidence in the
variation is given by the confidence percentile (for one interesting
parameter) of the statistical errors.  For instance, if $95\%$
confidence error bars do not overlap, but the $99\%$ errors do, then we
claim $\geq$$95\%$ confidence in the variation.  In general, we only
consider variations with $\geq$$90\%$ confidence to be significant.}
between observations.  Including Mrk~279 (the only Seyfert in which we
have looked for intra-observation variability), seven of sixteen
galaxies ($\sim$$45\%$) show significant variability of the Fe~K$\alpha$
line profile.  In $>$$50\%$ we also find variability of the line flux
and/or EW.  For most cases where we don't see variability, we are
hindered by poor photon counting statistics and sparse sampling.  For
instance, the four galaxies with only two pointings are also the only
four galaxies for which the EW doesn't vary with at least $68\%$
confidence.  If we only consider targets with at least four pointings,
we find that six out of nine exhibit variable line energies.

\subsection{Results for Individual Galaxies}

\subsubsection{Fairall~9}

This is one of the better-sampled galaxies, with a single 1993
observation followed a year later by a series of seven observations
spaced four days apart.  The source is variable, with the largest change
in 2--10~keV flux ($\sim$$30\%$) occurring in a four day span between
the second and third observations (Figure~8a).  The error bars are too
large to detect a corresponding change in the line flux or EW between
these two observations, but there are some general trends when
considering the entire data set.

The iron line energy changes significantly (at $99\%$ confidence),
with the largest change occurring between observations 7 and 8 (from
6.2 to 6.7~keV), again in a span of four days.  For these two
pointings, the line energy is lower when the source is fainter, but
there is no trend for such behavior when we compare all observations
(Figure~9a).

For the line flux and EW we plot both the $68\%$ and $90\%$ confidence
error bars in Figure~8a.  Neither the line flux nor the
EW changes at the $90\%$ confidence level, but they both do with
$>68\%$ confidence.  There is a hint that the EW is smaller in the
higher flux states which would be consistent with a time delay due to
a separation between the continuum source and the reprocessing
region.
If true, this implies a distance of at least 4 light days for the
reprocessor.

\subsubsection{MCG~8-11-11}

Between the two observations of MCG~8-11-11 (taken $<3$~days apart),
no change is observed in the 2--10~keV flux, line energy, or line flux
(Figure~8b).  The low EW reported for the 3 September 1995 observation
is a result of a large absorption correction.  When we measure the EW
without correcting for this absorption, we find a value of
$306\pm84$~eV, which is consistent with that obtained for the other
observation.

\subsubsection{NGC~3227}

The source flux changes by $\sim$$10\%$ between the two {\it ASCA}
pointings, approximately 2 years apart (Figure~8c).  No
line variability is observed, but there are only two time-averaged
data sets to be compared.  We note that during a single observation
the continuum changes by as much as a factor of two within a few
hours, so there may be interesting variability on time scales that are
comparable to Mkn~279.  (We have not yet studied in detail the
short-timescale behavior of this spectrum).

\subsubsection{NGC~3516}

NGC~3516 has been observed by {\it ASCA} on four occasions.  The source
flux varies by $\sim$$50\%$, decreasing most between the first and
second observations, separated by less than a year (Figure~8d).  The
line energy changes somewhat on long timescales.  Nandra et~al.\ (1999)
also find short timescale variability in the line profile. 

Changes in line flux and EW are significant.  The line flux follows the
continuum dropoff between observations 1 and 2 ($\Delta t \simeq 1$~yr),
but then increases again as the source becomes fainter ($\Delta t \simeq
3$~yr).  This latter change is accompanied by an increase in EW.  The
fact that the line first follows the continuum change and then becomes
stronger as the continuum weakens is difficult to understand in terms of
simple disk models of the iron line.

\subsubsection{NGC~3783}

Among the six observations of this source, the flux is seen to vary by
as much as $\sim$$70\%$ (Figure~8e).  The line energy changes by
$\sim$$3\%$, but there is no clear relationship between line energy and
flux, with the line having its highest energies during both low and
medium flux states.  The statistics are too poor to search for
variability in the line flux, but the line EW reaches its largest value
during the first observation, when the continuum is weakest.

\subsubsection{NGC~4051}

The 2--10~keV flux decreases by $\sim$$10\%$ between the two
{\it ASCA} pointings approximately one year apart
(Figure~8f).  We do not detect any changes in the 
iron line between the two observations, but the continuum varies by as 
much as a factor of $\sim$6 on timescales of a few hours and the 
short-term variability of the line has been investigated in 
detail by Wang et~al.\ (1999b).  Using a model that consists of a 
single power law and narrow Gaussian, these authors find that 
the line profile changes in ways that are independent of the 
continuum.  A very interesting result is that the EW of 
the Fe~K line in NGC~4051 increases as the source brightens
(Wang et~al.\ 1999b), as 
opposed to MCG~--6-30-15, where the EW decreases as the source 
brightens (this paper and Wang et~al.\ 1999b).

\subsubsection{NGC~4151}

Between the seven {\it ASCA} pointings, the 2--10~keV flux varies by as
much as $60\%$ (Figure~8g).  The line core energy remains constant
during this time.  The line flux also remains constant, although the
error bars are large.  Note that for this galaxy we discarded data below
2~keV to mitigate the complexities of the ionized absorber.

Most interesting is the trend for the EW to inversely correlate with the
continuum (Figure~9b).  The combination of constant line energy,
constant flux and decreasing EW with increasing source flux suggests
that the line core is dominated by emission far from the black hole.
From the time between observations 6 and 7, where the line has not
recovered its mean EW, we can place a lower limit of $\sim$1.4~years on
the lag in response of the line flux to the continuum, which corresponds
to an emission region larger than 0.43~pc ($8.8\times10^4$~M$_8^{-1} \;
\Rg$).  If we use the time scale of the full range of observations,
2.0~years, to define the minimum lag, the emission region is larger than
0.61~pc ($1.3 \times 10^5$~M$_8^{-1} \; \Rg$).

\subsubsection{NGC~4593 (Mrk~1330)}

The 2--10~keV flux increases by $25\%$ between the two {\it ASCA}
pointings, approximately 3.5 years apart (Figure~8h).
No changes are seen in the iron line between these two observations.

\subsubsection{MCG~--6-30-15}

The 2--10~keV flux varies by $40\%$ between the five {\it ASCA}
pointings (Figure~8i).  Changes in the broad line profile are well
documented for this galaxy (e.g., Iwasawa et~al.\ 1996 and Iwasawa
et~al.\ 1999).  Here, this change is reflected as a change in the
effective line energy.  We are insensitive to changes in the line on
short timescales, but on the longest timescales, we find behavior
similar to NGC~4151 in that the line core flux remains fairly constant
while the EW decreases with increasing source flux (Figure~9c, lower
panel).  However, unlike NGC~4151, which has a constant line energy,
MCG~--6-30-15 has a variable line energy, which suggests that processes
are occurring on small spatial scales.  The lack of reverberation
signatures on long timescales is consistent with the results of Reynolds
(2000) and Lee et~al.\ (1998), based primarily on {\it RXTE} data.

\subsubsection{IC~4329a}

This target has been observed on five occasions.  The 2--10~keV flux
changes by $\sim$$25\%$ (Figure~8j). The line energy changes by less
than $4\%$ but the data are not sensitive enough to comment on changes
in the EW or line flux.

\subsubsection{NGC~5548}

The 2--10~keV flux varies by $\sim$$50\%$ (Figure~8k).  When the iron
line is detected, its energy and flux remain constant.  However, the
Fe~K line is detected in only four of the eight observations.  To derive 
upper limits, we fix the line energy at its mean observed value.  In
these cases, we find that the line flux and EW has decreased
significantly (with $99\%$ confidence), in particular for observation~7
when the source is returning to its bright state.  (See line EW vs.\ 
continuum correlation in figure~9d).

We have examined whether the disappearance of the line can result from
our technique of fitting a narrow Gaussian profile to a line that has
become extremely broadened.  From a series of fits using a broad
Gaussian, both with and without the high-energy power-law component,
we find no evidence for a broad line for cases where the narrow line
has vanished.
 
\subsubsection{Mrk~841}

The 2--10~keV flux increases by $70\%$ between 1994 and 1997
(Figure~8l).  The iron line is not detected in the highest flux state
(the third of three observations) and the upper limit on the EW is
formally inconsistent with the EW in the first observation (with $95\%$
confidence).  This is similar to the behavior seen in NGC~5548.

\subsubsection{Mrk~509}

The 2--10~keV flux changes by as much as a factor of two (Figure~8m).
The iron line is detected in only five of eleven observations.  When
detected, its energy peaks near 6.4~keV, except for observation 10 where
the line is detected at 7~keV (6.4~keV is ruled out with $>95\%$
confidence).  The statistics are too poor to say much about the line
flux and EW, although both are significantly low during the first
observation of this galaxy in 1994, as well as in the non-detections of
the 2nd and 8th observations.

\subsubsection{NGC~7469 (Mrk~1514)}

Between the four {\it ASCA} pointings, the 2--10~keV flux varies by as
much as $50\%$ (Figure~8n).  The line energy changes but the errors on
line flux and EW are too large to comment more on the variability of
these parameters.  In a month-long {\it RXTE} monitoring campaign, the
line flux was found to be well correlated with the 2--10~keV continuum
flux, with evidence that the line emitting region lies within a few
light days of the continuum source (Nandra et~al.\ 2000).

\subsubsection{MCG~--2-58-22 (Mrk~926)}

MCG~--2-58-22 has been observed by {\it ASCA} three times.  The
2--10~keV flux increases by $100\%$ in 4 years (Figure~8o).  The iron
line is not detected in the second observation, and the upper limits on
the line flux and EW of this observation differ from the other two
measurements with $>95\%$ confidence.

\section{Discussion}

Our Fe~K results are summarized in Figure~\ref{fig:summary}.  By
examining all of the {\it ASCA} data for Seyfert~1 galaxies that have
multiple pointings available in the archive as of 1999 August, we have
shown that variability of the Fe~K$\alpha$ line profile on timescales of
days to years is common, occurring in at least $40\%$ of the galaxies.
We also observe changes in the line flux and/or EW in $\ga 50\%$ of the
galaxies.  In the one case where we have looked for variability during a
single observation (Mkn~279, which has only been observed once by
{\it ASCA}), we find variability of the line profile on a $\sim$10,000~s
timescale.

It is difficult to quantify for an individual galaxy how the changes
in the Fe~K$\alpha$ line are related to changes in the continuum.
This is primarily due to inadequate statistics and poor sampling.
However, interesting limits on the reprocessing material and black
hole mass can be derived for some cases.  Here we examine those limits
and discuss various physical mechanisms that may be responsible for
producing the observed variability.

\subsection{Line Variability in Mrk~279} 

Mkn~279 possesses an essentially featureless power law continuum, an
Fe~K$\alpha$ emission line with an equivalent width of $\sim$200~eV
that is possibly resolved, and evidence for Compton reflection.
Missing are low-energy absorption features such as O~VII and O~VIII
edges that are the signature of a warm absorber (Reynolds 1997), and
that are commonly seen in other Seyfert~1s.  At low energies the {\it
Rosat} PSPC spectrum is steeper than the {\it ASCA} spectrum by a
factor of $\Delta\Gamma \sim 0.4$.  This upturn suggests a distinct
and possibly variable soft X-ray emission component (see Appendix).

Using a narrow Gaussian to model the core of the Fe~K$\alpha$ line, we
find evidence for variability in the line profile.  As the continuum
source increases by an amplitude of $\sim$$20\%$ in 10~ks, the peak
energy of the iron line changes by $\sim$$3\%$ ($\sim$10,500~km~s$^{-1}$; 
Figure~\ref{fig:mergedCont}).  The rapid response of the Fe~K$\alpha$
line to the continuum suggests an emission region that is less than a few
light-hours across.  The light crossing time on a region of the disk with
a radius of $x$ in \Rg\ units is $x \; \Rg /c \sim 500x \; {\mathrm M}_8$~s.
If we assume the Fe~K line is emitted by material in the disk which is
orbiting at a distance of at least the marginally stable orbit
(in the Schwarzschild metric; $\mathrm{r_{ms}} = 6 \Rg$), this implies
a central black hole mass $\la 3 \times 10^8 \; \Msun$.

\subsection{Line Variability in the Seyfert~1 sample}

In $40\%$ of Seyfert~1s with multiple {\it ASCA} observations, we detect
variability of the iron line profile on timescales of days to years, as
measured by a change in the effective line peak energy
(Figure~\ref{fig:summary}).  The detection rate jumps to $67\%$ (6 out
of 9) if we only consider the galaxies with better sampling.  By
examining only the line cores, our measurement represents a
{\it lower limit} to the variability and will undersample changes in the
entire line profile.  Variability in the core of the line is interesting
however, because the core is expected to be the least variable part of
the line if it originates far from the black hole, as is the case in
Seyfert~2 galaxies like NGC~2992 (Weaver et~al.\ 1995).

The temporal behavior of the Fe~K$\alpha$ line is complex, and sparse
sampling hinders our ability to detect any clear variability patterns.
The two cases which provide enough information to establish a trend are
NGC~4151 and MCG~--6-30-15.  For NGC~4151, the line flux remains
relatively constant on timescales of months to years and does not
respond to changes in the continuum.  This behavior falls in line with
all other historical measurements, which are consistent with a constant
line intensity despite large continuum changes.   The lag in response of
the Fe~K line to the continuum suggests that the line originates far
from the source of illumination, a conclusion also supported by the fact
that the peak energy of the line core stays constant throughout the
{\it ASCA} observations.  Other, more indirect evidence for a partly
large emission region is the complex line profile, with a narrow,
centrally-peaked component at 6.4~keV superimposed on a broad component
(Wang et~al.\ 1999a, Yaqoob et~al.\ 1995).  The line properties are thus
consistent with simple reverberation models including a non-disk
contribution to the emission, possibly from the obscuring torus.
{\it BBXRT} observed a lack of flux variability in the Fe~K line for a
17 hour span, leading to an estimate of $5.8\times10^{-5}$~pc ($= 37 \;
{\mathrm M}_8^{-1} \Rg$) as the shortest distance from the central
source to the region where the line originates (Weaver et~al.\ 1992).
{\it ASCA} observations have now increased this limit to 2.0~years or
0.61~pc, which corresponds to $1.3 \times 10^5 \; {\mathrm M}_8^{-1} \Rg$.

For MCG~--6-30-15, our results show that the line flux remains
approximately constant between observations and the EW is inversely
proportional to the brightness of the source.  However, unlike NGC~4151,
the peak energy of the line changes on long timescales (this paper) and
short timescales (Iwasawa et~al.\ 1996).  The short-term changes in the
line profile imply a small emission region.  The temporal/spectral
behavior of the line is difficult to reconcile with simple models of an
accretion disk that produces the Fe~K emission from regions close to the
black hole.  The light crossing time of the inner disk is short compared
to the timescales of variability we have examined for sources in the
sample, yet we do not generally observe the line flux to respond to the
continuum variability as expected.  Whether the X-ray source is situated
at a central location above the accretion disk (the so-called ``lamppost
model'', e.g. Nayakshin \& Kallman 2000) or in the form of a hot corona
on the disk surface (the so-called ``sandwich model'', e.g. Collin
et~al.\ 2000), one would expect the line flux to track the continuum,
yielding a constant equivalent width over the time scales studied, as
well as possible changes in the line profile.  On the other hand, for a
region as large as the outer disk or torus, the light crossing time
would be greater than the temporal resolution of the data, in which case
we expect the line flux to lag the continuum and we don't expect changes
in the line profile.  Obviously, the Fe~K line, with it's variable
profile {\it and} constant flux, does not easily fit either scenario.

Other objects in our sample possess unusual behavior.  In NGC~3516, the
line seems to vary independently of the continuum, with the flux
dropping with the continuum over the span of a year but then appearing
to strengthen again as the continuum drops further in a three-year span.
Similar behavior occurs for this galaxy on short timescales (Nandra
et~al.\ 1999).  In NGC~5548, the line is undetected in half of the
observations with fairly strict upper limits, but whether or not the
line is present does not seem to relate to the continuum flux state.
For Mkn~509, the Fe~K line energy increases to $\sim$7~keV during one
observation, but this increase does not appear to be related to a change
in the continuum.

For the lamppost model, changes in line energy can be explained by
changes in the ionization of the disk for different flux states.
However, the behavior of the line profile appears to bear little
relation to the continuum in general, and so there must be something
else responsible for controlling the line shape and/or intensity.  One
possibility is a physical change in the reprocessor.  If the geometry of
the accretion disk changes, with the disk extending farther toward the
black hole in certain flux states, for example, then this could result
in enhanced Doppler shifts resulting in a more significant
Doppler-boosted peak and redshifted wing.  Another possibility is some
kind of time-variable, non-axisymmetric obscuration (Weaver \& Yaqoob
1998), which can remove photons in the line core, shifting the mean
energy.  On the other hand, short-term changes could result from
localized flaring phenomena, such that the X-ray illumination is spread
over the disk surface unevenly (Nayakshin \& Kallman 2000) and most of
the X-ray reprocessing happens near the flare locations (e.g., the
rotating hot-spot model; Ruszkowski 2000).

Observations with {\it XMM} and {\it Constellation-X} will be crucial
to understand the variability patterns of the iron K lines.  However,
even with very high-quality data, interpreting time-resolved spectra
will require assumptions regarding the accretion disk structure, disk
instabilities, diffusion rates, and disk winds and shocks.  A roadmap
for understanding these types of observations already exists in other
disciplines such as studying disk emission in galactic microquasars
(Markwardt, Swank \& Taam 1999; Eikenberry et~al.\ 1998), but such
techniques have yet to be applied to the large accreting systems in
active galaxies.  It will be important to see if it is possible to
relate changes in the lines to specific changes in ionization state,
changes in illumination, or changes in disk structure.  It may be
difficult to distinguish between basic Doppler, gravitational, and
ionization effects, and the movement of localized X-ray/UV hot spots
until we can image the accretion disk in X-rays.

\section{Summary and Conclusions}

We have studied variability of the Fe~K$\alpha$ line on short time
scales in one source, and on longer time scales in a sample of Seyfert~1
galaxies.  In the {\it ASCA} observation of Mkn~279 we saw a $20\%$
increase in 2--10~keV flux over $\sim$10~ks.  In this time, the energy
of the Fe~K$\alpha$ line core increases from 6.30$^{+0.10}_{-0.12}$~keV
in the low state to 6.52$^{+0.09}_{-0.09}$~keV in the high state.

The procedure developed for Mkn~279 has allowed us to study the line
variability across an entire {\em sample} of Seyfert galaxies.  By
examining the time-averaged data for the 15 Seyfert~1 galaxies with
multiple {\it ASCA} pointings, we find that variability of the
Fe~K$\alpha$ line profile is common, appearing in $40\%$ of the objects,
and in two thirds of those galaxies which have been observed at least
four times.  Variability in the line flux and/or EW occur in
$\sim$$50\%$ of the galaxies.  Because we have examined only the time
averages of the line cores, and also due to the limited sampling, our
measurements represent a {\it lower limit} to the true variability of
the line.

The general behavior of the line across the sample bears little relation
to the continuum.  The data clearly show that in no galaxy does the
Fe~K$\alpha$ line simply track the continuum.  Beyond this, limited
sampling makes it difficult to quantify how changes in the Fe~K$\alpha$
line are related to changes in the continuum for most galaxies.  In
NGC~4151 the line core appears to be constant, independent of continuum
changes on time scales up to two years; for other Seyfert~1s the Fe line
behavior is more complex, and is generally difficult to explain in the
context of our current theoretical models.  Taken as a whole, we have
confirmed with the largest line variability study to date that the
simple disk model is inadequate to explain the data.  We need to
reexamine the assumptions of our models from a theoretical perspective
in order to refine them and make them consistent with observations.

\bigskip
\acknowledgments

We thank Ken Ebisawa, Keith Gendreau and Koji Mukai for helpful
discussions about the {\it ASCA} calibration effort and its current
status.

% \clearpage 
% \vfill\eject

\appendix
\section{Appendix: Concerns Regarding {\it ASCA} Calibration}

\noindent
There have been ongoing discussions about calibration issues involving 
the SIS detector response at low energies ($< 1$~keV).  These
calibration problems are especially noticeable when attempting joint
spectral analysis with the {\it Rosat} PSPC.  Part of the problem with
joint fits appears to lie with the PSPC, which is known to have
temporally-dependent energy nonlinearity and residual,
temporally-dependent spatial gain variations (Snowden, Turner \&
Freyberg 2000).  However, part of the problem also stems from the fact
that the SIS calibration has not kept pace with the in-flight
instrument degradation.  There is evidence that the SIS detector
response has been degrading steadily since launch, not only in energy
resolution but also in loss of effective area at low energies due to
the time-dependent effects of radiation damage.

\noindent
Current software does not account for all of these changes.  One
example of discrepant results in the literature is a study of X-ray
bright galaxy groups by Hwang et~al.\ (1999).  Comparing {\it ASCA}
and the PSPC, these authors find differences between the thermal
temperatures and column densities in the sense that the SIS
temperatures are systematically lower than the PSPC temperatures and
the SIS column densities are systematically higher than the PSPC
column densities (the latter tend to agree with Galactic values, as
expected).  Another example is the Seyfert galaxy NGC~5548, for which
simultaneous {\it Rosat} and {\it ASCA} observations are available.
For this galaxy, Iwasawa, Fabian and Nandra (1999) find a clear and
unexplained mismatch of photon indices ($\Delta\Gamma \simeq 0.4$)
between the SIS and the PSPC.  We cite Mkn~279 as another example.
The {\it ASCA} photon index is significantly less than the PSPC index
($\Delta\Gamma \simeq 0.4$)\footnote{A smaller index
in Mkn~279 also shows up in the GIS and so the larger index in the
{\it Rosat} band may be real.} and the SIS1 column density is
significantly larger than the PSPC column density.

\noindent
In this appendix we examine the nature of the SIS calibration problem, 
estimate its magnitude and suggest ways to cope with it at the
analysis stage.  We attempt to explain the {\it ASCA} SIS quantum
efficiency (``QE'') problem in a self-consistent manner without
relying on cross-measurements with {\it Rosat}.  As we have already
stated, the miscalibration between {\it ASCA} and {\it Rosat} is not
totally due to {\it ASCA}, but the {\it ASCA} problem needs to be
quantified.

\noindent
There are a number of ways to examine the calibration; some being more 
model-dependent than others.  A rigorous and model-independent
approach based on examining the ratio of data from the various
detectors is followed by Yaqoob in a recent calibration memo
(Yaqoob et~al.\ 2000).  Our method is based on model fitting, but its
strength lies in using the GIS as a cross-calibrator for the SIS.  We
find that the GIS is sensitive enough to test for small {\it
differences} in \NH\ on the order of a few $\times10^{20} \;
\mathrm{cm}^{-2}$, although the  {\it absolute} calibration is not as
accurate as this.

\subsection{Cross-calibration of the SIS0 and SIS1}

\noindent
Very briefly, the {\it ASCA} detectors were calibrated along the
following lines.  The GIS was calibrated using the Crab, which is too
bright for the SIS.  Therefore, the SIS had to calibrated against the
GIS.  SIS data for the quasar 3C~273 between 2--6~keV were then used
to adjust the depletion depth of the SIS model to obtain agreement
with GIS data for the same target.  The final adjustment of the QE
curve for the SIS was released in 1994 November (version 0.8 of sisrmg
and the response matrices), and was based on an observation of 3C~273
taken on 16 December 1993.  At this point the QE was adjusted again
relative to the ground calibration to force consistency between the
four {\it ASCA} instruments.  At the time, it was not realized that
the discrepancy that was being corrected was changing with time and so
the baseline QE has become out of date.  Since then, the response
matrix generator has changed in substance only to model certain
energy-resolution changes in the detectors.

\noindent
The degradation of the SIS detectors due to the on-orbit effects of
radiation damage on the CCDs shows up as an increasing loss of photons
below 1~keV with time.  This is manifestly different from the
%so-called `RDD' (Residual Dark current Distribution) problem (even
so-called ``RDD'' (Residual Dark current Distribution) problem (even
then, RDD correction software only mitigates some of the RDD
problems).  The reasons for the additional losses are not understood
by the calibration teams.  It is suspected that it is due to
increasing dark current and loss of Charge Transfer Efficiency (CTE).
The problem is still under investigation by the instrument teams.  The
current trend is more pronounced in S1 than S0 and shows up in 1, 2,
and 4~CCD mode data.  The absolute magnitude of the effect is larger
for modes using a larger number of CCDs but curiously, the {\it ratio}
of the loss of efficiency seems to be independent of CCD mode (Yaqoob
et~al.\ 2000).

\noindent
The loss in SIS detector efficiency can be parameterized crudely as
excess absorption.  So, when using the current calibration,
observations made before 16 December 1993 have SIS data which appear
to turn up at low energies (a negative ``absorption'') while
observations made after 16 December 1993 have SIS data which appear
to turn down at low energies (a positive ``absorption,'' which has
increased with time).  December 16 1993 can therefore be viewed as an
anchor point at which the time-dependent correction investigated by us
and by Yaqoob et~al.\ (2000) is zero.

\noindent
Adopting the ``excess-\NH '' approach, we illustrate the effect by
first showing the magnitude of the effect for the {\it ASCA}
observation of Mrk~279 taken in December 1994.  Mrk~279 is an
excellent target to search for charge loss due to its naturally small
absorbing column.  Then we compare the SIS and GIS data for a sample
of AGN that were observed multiple times throughout the mission to
attempt a crude cross-calibration.

\paragraph{Mrk~279}  This galaxy has no intrinsic absorption according
to the PSPC and a fairly low Galactic column density, which makes it
an excellent target for soft X-ray calibration.   From joint fits with
the {\it ASCA} PSPC data, we have already noted that the photon
indices do not match, but here we compare the column densities.  For a
power-law model including a low-temperature thermal component to model
the steeper {\it Rosat} spectrum, we find that the PSPC and S0 measure
the same Galactic column density to within statistical uncertainties
(Figure~\ref{fig:contours}).  However, for the same model for S1,
\NH(S1) is larger by about $3\times10^{20}$~cm$^{-2}$ (for the same
photon index).  The consistency of column densities between S0 and the
PSPC agrees with the overall trend with time.  Mrk~279 was observed
early in the mission at a point when S0 had not degraded appreciably
while S1 had begun to degrade.

\noindent
\paragraph{The SIS and GIS}
Taking advantage of Seyfert~1 galaxies as cross-calibration sources
and comparing the SIS against the GIS, we examine trends for changes
in the apparent column density for observations taken in 1-CCD and
%mixing 1 and 2 ccd mode?
2-CCD mode.  We fit the data from 0.6 to 4~keV, forcing all of the
model parameters to be the same except for the column density.  \NH\ 
is set to be equal for the two GIS spectra but allowed to be a free
parameter for S0 and S1 independently.  Our results are summarized in
Figure~\ref{fig:extraNh}.  For S0, we see a systematic trend for
\NH(S0) to increase approximately linearly with time since launch,
from $\sim$$2\times10^{20}$~cm$^{-2}$ in 1995 to
$\sim$$6\times10^{20}$ in 1998.   The S0 data fit very well with a
straight line described by
\begin{displaymath}
\NH \mathrm{(SIS0)} = 3.635857 \times 10^{-8} 
	(T - 3.017482 \times 10^{7}) \times 10^{20}\;\mathrm{cm}^{-2}
\end{displaymath}
where $T$ is ASCATIME (average of the start and stop times of the
observation) in seconds since 1 January 1993.

\noindent
S1 behaves differently from S0.  The trend is not linear, but peaks at 
around 1500 days since launch.  The S1/S0 discrepancy therefore shows
%no clear trend with time.  To `correct' for the change in S1, one
no clear trend with time.  To ``correct'' for the change in S1, one
could measure the ratio of S1/S0 for a given observation, determine
the excess NH(S1/S0),and then add this to NH(S0).  If the
signal-to-noise is not good enough to measure the S1/S0 ratio then one
should not be concerned about correcting the data since statistical
errors will dominate over systematic errors.

\noindent
We note that the GIS has an effect below 1~keV that mimics a {\it
negative} absorption of $\sim$$2\times10^{20}$~cm$^{-2}$ (the opposite
of the SIS).  The root cause for this discrepancy seems to be that the
%added that the SOFT response was independent
{\it soft} X-ray GIS and SIS responses were originally calibrated
independently.  The GIS/XRT response was tuned to the Crab to give a
column density of $3\times10^{21}$~cm$^{-2}$ while the SIS low-energy
efficiency was only calibrated on the ground.  The column density
assumed for the Crab carries a larger uncertainty than
$2\times10^{20}$~cm$^{-2}$ so the GIS calibration was not intended to
be good enough to measure {\it absolute} absorption to this level.
However, we believe that the GIS detectors are sensitive to {\it
differences} of this order of column density, so although they can't
be used to make absolute measurements of small column densities, they
can measure a small {\it relative} difference in \NH\ compared to the
SIS.

\noindent
The problem with the low-energy inflight calibration of the SIS is
that there is no suitable astrophysical source.  The ideal requirement 
is that such a source should be bright, have constant intensity, low
intrinsic absorption, a simple continuum, and it should be a point
source.  No such source exists.  3C~273 is a bright point source but
it is variable in both intensity and spectral shape.  There are also
remaining known problems with the GIS, most importantly uncertainty in
the energy to pulse-height relation and this problem is currently
being addressed by the instrument teams.  This will affect the energy
scale and have minor effects on the spectral response.  There will
also be refinements in the Beryllium window transmission which may
affect absolute column density measurements.

\noindent
The excess-\NH\ problem affects joint model fits of the PSPC and {\it
ASCA} if left uncorrected.  A global correction of S0 based on
Figure~\ref{fig:extraNh} may be adequate for sources with simple
spectra and moderate count rates, depending on science goals.  For
high resolution, high-quality spectra of complex line-emitting plasmas 
%this `correction' must be applied with caution because we do not
this ``correction'' must be applied with caution because we do not
understand the detailed changes of the effective area with energy and
it does not model any changes in energy resolution.

\noindent
\paragraph{Additional effects?}
For Mkn~279, the {\it ASCA} photon index is smaller than the PSPC
index even when we restrict the {\it ASCA} fits to energies above
1~keV, where {\it ASCA} is well calibrated.  So, is there a slope
problem different from the low-energy `cutoff' problem?  To determine
the magnitude of any additional error, we have examined PSPC and
{\it ASCA} data for MCG~--2-58-22, another bright Seyfert~1 galaxy
with no intrinsic absorption that was observed early in the mission
(Weaver et~al.\ 1995).  Including the `\NH\ correction' for
MCG~--2-58-22, we find that the PSPC and ASCA indices differ by less
than $\Delta\Gamma\sim0.1$.  This remaining slope offset is small
enough to suggest that the large slope offset for Mkn~279 is real, at
least in this case.

\vfill\eject

\begin{deluxetable}{lcccccccc}
%\rotate
\tablecolumns{10}
\tablewidth{0pc}
\tablecaption{{\it ASCA} Time-Averaged Spectral Results for
	Mkn~279\label{tab:integrated279}}
\tablehead{
Model\tablenotemark{a} & $\Gamma$ & E$_{\rm Fe~K\alpha}$ & 
   $\sigma_{\rm Fe~K\alpha}$ & EW$_{\rm Fe~K\alpha}$ & $\theta$ &
   $R$  & $\chi^2/dof$ \\
 & & [keV] & [keV] & [eV] & [$^{\circ}$] & & }
\startdata
1) PL      & $2.01^{+0.01}_{-0.01}$ & \nix                    
     & \nix                   & \nix              & \nix         
        & \nix                   & 922.9/860 \\
2) PL+Gn   & $2.02^{+0.01}_{-0.01}$ & $6.47^{+0.05}_{-0.06}$  
     & 0.01(f)                & $196^{+42}_{-42}$ & \nix         
        & \nix                   & 865.4/858 \\
3) PL+Gb   & $2.04^{+0.03}_{-0.03}$ & $6.50^{+0.12}_{-0.17}$  
     & $0.51^{+1.39}_{-0.27}$ & $440^{+90}_{-80}$ & \nix         
        & \nix                   & 846.6/857 \\ 
4) PL+Dl   & $2.03^{+0.02}_{-0.02}$ & $6.50^{+0.08}_{-0.08}$  
     & \nix                   & $560^{+80}_{-80}$ & $89_{-55}^{*}$
        & \nix                   & 841.1/857 \\ 
5) PL+R    & $2.11^{+0.03}_{-0.03}$ & \nix                    
     & \nix                   & \nix              & 30(f)        
        & $2.54^{+0.63}_{-0.57}$ & 848.5/859 \\
6) PL+Gn+R & $2.10^{+0.02}_{-0.02}$ & $6.47^{+0.07}_{-0.07}$  
     & .01(f)                 & $120^{+45}_{-45}$ & 30(f)        
        & $1.91^{+0.62}_{-0.58}$ & 825.8/857 \\
7) PL+Gb+R & $2.10^{+0.02}_{-0.02}$ & $6.46^{+0.09}_{-0.09}$  
     & $0.14^{+0.14}_{-0.14}$ & $148^{+68}_{-68}$ & 30(f)        
        & $1.60^{+0.50}_{-0.62}$ & 824.3/856 \\
8) PL+Gb+R & $2.06^{+0.02}_{-0.02}$ & $6.46^{+0.09}_{-0.09}$  
     & $0.21^{+0.19}_{-0.11}$ & $213^{+65}_{-65}$ & 30(f)        
        & 1.0(f)                 & 829.8/857 \\
9) PL+Dl+R & $2.09^{+0.02}_{-0.02}$ & $6.46^{+0.10}_{-0.10}$
     & \nix                   & $210^{+70}_{-70}$ & $15_{-15}^{+35}$
        & $1.50^{+0.44}_{-0.53}$ & 825.9/856 \\
\enddata

\tablenotetext{a} {Model components are PL:~a power law with a
normalization of
$1.03\pm0.02\times10^{-2}$~photons~keV$^{-1}$~cm$^{-2}$~s$^{-1}$ at
1~keV, Gn:~narrow (unresolved) Gaussian, Gb:~broad Gaussian, Dl:~disk
line, and R:~Compton reflection.  For the line we use the
Schwarzschild disk-line model with $\mathrm{R_{in}} = 6 \; \Rg$,
$\mathrm{R_{out}} = 1000 \; \Rg$, and an emissivity index $q = 2.5$,
where emissivity $\propto \mathrm{r}^{-q}$.
}
\tablecomments{Includes grade 6 photon events from the SIS.  Fixed
parameters are labeled by (f).  All models include absorption by the
Galaxy ($\NH\ = 1.64 \times 10^{20}$~cm$^{-2}$). Data are fitted from
1 to 10~keV.  Errors represent $90\%$ significance.  The columns are: 
(1)~model label and description, (2)~photon index, (3)~rest frame line
energy, (4)~line width, (5)~line equivalent width, (6)~inclination of
the accretion disk, (7)~reflection normalization, and (8)~fit quality.
}
\end{deluxetable}

\clearpage 

% \begin{deluxetable}{lccccccccc}
% \begin{deluxetable}{l@{\ \ }c@{\ \ \ }c@{}c@{\ }c@{\ \ \ }c@{\ \ \ }c@{\ \ \ }c@{\ \ }c@{\ \ }c}
\begin{deluxetable}{l@{\ \ }c@{\ \ }c@{}c@{\ }c@{\ \ }c@{\ \ }c@{\ \ }c@{\ }c@{\ }c}
% \tabletypesize{\small}
% \rotate
\tablecolumns{10}
\tablewidth{0pc}
\tablecaption{Line Properties for a Narrow Gaussian\label{tab:sample}}
\tablehead{
Galaxy & sequence & Date & CCD & Time & $\mathrm{E_{K \alpha}}$ 
  & $\mathrm{N_{K \alpha}}$ & $\mathrm{EW_{K \alpha}}$  
     & $\Delta\chi^2$ & $\mathrm{F_{2-10}}$ \\ 
   &  &  & mode & [ks] & [keV] & [$10^{-5}$] & [eV] &  & [$\times 10^{-12}$] }
\startdata
Fairall~9     & 71027000 &           21 Nov 93 & 1 & 22.0
  & $6.32^{+0.07}_{-0.07}$ & $\phantom{1}3.1^{+0.9}_{-0.9}$
  & $          120^{+ 36}_{- 35}$ & \phantom{1}26 & $22.0^{+0.8}_{-0.9}$ \\*
Fairall~9     & 73011000 & \phantom{1}2 Dec 94 & 1 & 16.9
  & $6.33^{+0.08}_{-0.09}$ & $\phantom{1}3.1^{+1.3}_{-1.3}$
  & $\phantom{1}92^{+ 40}_{- 40}$ & \phantom{1}12 & $27.8^{+1.1}_{-1.3}$ \\*
Fairall~9     & 73011010 & \phantom{1}6 Dec 94 & 1 & 15.2
  & $6.52^{+0.12}_{-0.13}$ & $\phantom{1}3.1^{+1.3}_{-1.3}$
  & $          128^{+ 55}_{- 55}$ & \phantom{11}9 & $21.5^{+1.3}_{-1.1}$ \\*
Fairall~9     & 73011020 &           10 Dec 94 & 1 & 18.3
  & $6.47^{+0.11}_{-0.09}$ & $\phantom{1}4.4^{+1.3}_{-1.3}$
  & $          171^{+ 51}_{- 50}$ & \phantom{1}25 & $24.7^{+1.1}_{-1.1}$ \\*
Fairall~9     & 73011030 &           14 Dec 94 & 1 & 18.1
  & $6.35^{+0.09}_{-0.10}$ & $\phantom{1}2.9^{+1.3}_{-1.3}$
  & $\phantom{1}85^{+ 38}_{- 38}$ & \phantom{1}11 & $24.6^{+1.2}_{-1.0}$ \\*
Fairall~9     & 73011040 &           18 Dec 94 & 1 & 21.4
  & $6.35^{+0.08}_{-0.08}$ & $\phantom{1}3.8^{+1.2}_{-1.2}$
  & $          122^{+ 39}_{- 39}$ & \phantom{1}21 & $27.6^{+1.0}_{-1.2}$ \\*
Fairall~9     & 73011060 &           22 Dec 94 & 1 & 16.4
  & $6.65^{+0.12}_{-0.15}$ & $\phantom{1}2.6^{+1.4}_{-1.4}$
  & $\phantom{1}84^{+ 47}_{- 47}$ & \phantom{11}6 & $26.1^{+1.1}_{-1.3}$ \\*
Fairall~9     & 73011050 &           26 Dec 94 & 1 & 7.38
  & $6.20^{+0.09}_{-0.10}$ & $\phantom{1}4.1^{+1.7}_{-1.7}$
  & $          158^{+ 67}_{- 66}$ & \phantom{1}12 & $22.3^{+1.4}_{-1.5}$ \\
MCG~8-11-11\tablenotemark{a}
              & 73052000 & \phantom{1}3 Sep 95 & 1 & 11.5
  & $6.45^{+0.05}_{-0.06}$ & $\phantom{1}7.2^{+2.0}_{-2.0}$
  & $          7.4^{+2.1}_{-2.1}$ & \phantom{1}23 & $26.9^{+1.7}_{-1.5}$ \\*
MCG~8-11-11   & 73052010 & \phantom{1}6 Sep 95 & 1 & 11.0
  & $6.45^{+0.09}_{-0.07}$ & $\phantom{1}6.0^{+2.1}_{-2.1}$
  & $          179^{+ 63}_{- 63}$ & \phantom{1}17 & $25.3^{+1.8}_{-1.5}$ \\
NGC~3227      & 70013000 & \phantom{1}8 May 93 & 4 & 36.4
  & $6.42^{+0.04}_{-0.04}$ & $\phantom{1}4.7^{+1.0}_{-1.0}$
  & $          127^{+ 27}_{- 27}$ & \phantom{1}48 & $35.8^{+1.0}_{-1.0}$ \\*
NGC~3227      & 73068000 &           15 May 95 & 1 & 35.8
  & $6.36^{+0.07}_{-0.07}$ & $\phantom{1}5.7^{+1.2}_{-1.2}$
  & $          146^{+ 30}_{- 29}$ & \phantom{1}62 & $31.5^{+1.1}_{-1.1}$ \\
NGC~3516      & 71007000 & \phantom{1}2 Apr 94 & 1 & 30.1
  & $6.40^{+0.04}_{-0.04}$ & $          11.0^{+1.8}_{-1.8}$
  & $          122^{+ 20}_{- 20}$ & \phantom{1}95 & $86.5^{+1.6}_{-1.6}$ \\*
NGC~3516      & 73066000 &           11 Mar 95 & 1 & 19.9
  & $6.38^{+0.06}_{-0.07}$ & $\phantom{1}5.6^{+1.8}_{-1.8}$
  & $\phantom{1}94^{+ 29}_{- 29}$ & \phantom{1}23 & $54.5^{+1.6}_{-1.6}$ \\*
NGC~3516      & 73066010 &           12 Mar 95 & 1 & 19.0
  & $6.33^{+0.06}_{-0.06}$ & $\phantom{1}7.4^{+1.8}_{-1.8}$
  & $          124^{+ 30}_{- 29}$ & \phantom{1}40 & $54.0^{+1.6}_{-1.5}$ \\*
NGC~3516      & 76028000 &           12 Apr 98 & 1 & 153.
  & $6.33^{+0.02}_{-0.02}$ & $\phantom{1}8.5^{+0.7}_{-0.7}$
  & $          149^{+ 11}_{- 12}$ &           396 & $47.6^{+0.6}_{-0.6}$ \\
NGC~3783      & 71041000 &           19 Dec 93 & 2 & 17.5
  & $6.39^{+0.04}_{-0.04}$ & $          10.4^{+2.0}_{-2.0}$
  & $          164^{+ 32}_{- 32}$ & \phantom{1}67 & $50.0^{+1.7}_{-1.7}$ \\*
NGC~3783      & 71041010 &           23 Dec 93 & 2 & 16.2
  & $6.24^{+0.07}_{-0.06}$ & $\phantom{1}7.6^{+2.3}_{-2.3}$
  & $\phantom{1}95^{+ 29}_{- 28}$ & \phantom{1}29 & $66.3^{+2.1}_{-2.0}$ \\*
NGC~3783      & 74054000 & \phantom{1}9 Jul 96 & 1 & 17.6
  & $6.26^{+0.07}_{-0.08}$ & $\phantom{1}8.5^{+2.5}_{-2.5}$
  & $\phantom{1}88^{+ 26}_{- 26}$ & \phantom{1}30 & $85.8^{+2.3}_{-2.2}$ \\*
NGC~3783      & 74054010 &           11 Jul 96 & 1 & 20.4
  & $6.34^{+0.07}_{-0.07}$ & $\phantom{1}8.1^{+2.3}_{-2.3}$
  & $\phantom{1}93^{+ 27}_{- 27}$ & \phantom{1}30 & $78.0^{+2.0}_{-2.0}$ \\*
NGC~3783      & 74054020 &           14 Jul 96 & 1 & 18.0
  & $6.35^{+0.06}_{-0.06}$ & $          10.3^{+2.5}_{-2.5}$
  & $          117^{+ 29}_{- 28}$ & \phantom{1}42 & $82.8^{+2.2}_{-2.2}$ \\*
NGC~3783      & 74054030 &           16 Jul 96 & 1 & 17.7
  & $6.38^{+0.04}_{-0.04}$ & $          10.8^{+2.5}_{-2.5}$
  & $          124^{+ 30}_{- 29}$ & \phantom{1}45 & $77.9^{+2.2}_{-2.1}$ \\
NGC~4051      & 70001000 &           25 Apr 93 & 4 & 27.2
  & $6.43^{+0.06}_{-0.06}$ & $\phantom{1}3.5^{+1.0}_{-1.0}$
  & $          134^{+ 39}_{- 39}$ & \phantom{1}27 & $27.7^{+0.9}_{-0.9}$ \\*
NGC~4051      & 72001000 & \phantom{1}7 Jun 94 & 1 & 72.0
  & $6.38^{+0.05}_{-0.05}$ & $\phantom{1}2.7^{+0.6}_{-0.6}$
  & $          100^{+ 24}_{- 24}$ & \phantom{1}41 & $25.6^{+0.5}_{-0.6}$ \\
NGC~4151\tablenotemark{b}
              & 70000000 &           24 May 93 & 4 & 13.6
  & $6.38^{+0.03}_{-0.03}$ & $          37.2^{+5.0}_{-5.0}$
  & $          108^{+ 15}_{- 14}$ &           122 & $167.^{+6. }_{-6. }$ \\*
NGC~4151\tablenotemark{b}
              & 70000010 & \phantom{1}5 Nov 93 & 1 & 7.26
  & $6.35^{+0.04}_{-0.04}$ & $          31.4^{+7.1}_{-7.1}$
  & $\phantom{1}73^{+ 16}_{- 16}$ & \phantom{1}39 & $222.^{+6. }_{-5. }$ \\*
NGC~4151\tablenotemark{b}
              & 71019030 & \phantom{1}4 Dec 93 & 2 & 11.6
  & $6.35^{+0.04}_{-0.04}$ & $          33.7^{+6.5}_{-6.5}$
  & $\phantom{1}72^{+ 14}_{- 14}$ & \phantom{1}61 & $263.^{+7. }_{-6. }$ \\*
NGC~4151\tablenotemark{b}
              & 71019020 & \phantom{1}5 Dec 93 & 2 & 10.9
  & $6.36^{+0.04}_{-0.04}$ & $          32.3^{+6.4}_{-6.4}$
  & $\phantom{1}76^{+ 15}_{- 15}$ & \phantom{1}56 & $232.^{+6. }_{-6. }$ \\*
NGC~4151\tablenotemark{b}
              & 71019010 & \phantom{1}7 Dec 93 & 2 & 12.4
  & $6.38^{+0.03}_{-0.02}$ & $          35.3^{+6.3}_{-6.3}$
  & $\phantom{1}77^{+ 14}_{- 14}$ & \phantom{1}71 & $268.^{+6. }_{-6. }$ \\*
NGC~4151\tablenotemark{b}
              & 71019000 & \phantom{1}9 Dec 93 & 2 & 12.2
  & $6.36^{+0.03}_{-0.03}$ & $          39.0^{+6.1}_{-6.1}$
  & $\phantom{1}90^{+ 14}_{- 14}$ & \phantom{1}88 & $255.^{+6. }_{-6. }$ \\*
NGC~4151\tablenotemark{b}
              & 73019000 &           10 May 95 & 1 & 96.8
  & $6.37^{+0.01}_{-0.01}$ & $          35.7^{+1.9}_{-1.9}$
  & $          101^{+  5}_{-  5}$ &           777 & $180.^{+2. }_{-2. }$ \\
NGC~4593      & 71024000 & \phantom{1}9 Jan 94 & 1 & 22.1
  & $6.33^{+0.08}_{-0.07}$ & $\phantom{1}3.7^{+1.3}_{-1.3}$
  & $\phantom{1}70^{+ 25}_{- 25}$ & \phantom{1}17 & $40.4^{+1.0}_{-1.3}$ \\*
NGC~4593      & 75023010 &           20 Jul 97 & 1 & 14.0
  & $6.39^{+0.08}_{-0.08}$ & $\phantom{1}5.2^{+2.1}_{-2.1}$
  & $\phantom{1}83^{+ 34}_{- 34}$ &           105 & $50.8^{+1.8}_{-1.7}$ \\
MCG~--6-30-15 & 70016000 & \phantom{1}9 Jul 93 & 2 & 30.7
  & $6.27^{+0.07}_{-0.07}$ & $\phantom{1}3.3^{+1.3}_{-1.3}$
  & $\phantom{1}58^{+ 23}_{- 23}$ & \phantom{1}14 & $52.0^{+1.3}_{-1.2}$ \\*
MCG~--6-30-15 & 70016010 &           31 Jul 93 & 1 & 29.3
  & $6.40^{+0.05}_{-0.05}$ & $\phantom{1}4.4^{+1.3}_{-1.3}$
  & $          102^{+ 31}_{- 31}$ & \phantom{1}27 & $42.6^{+1.2}_{-1.2}$ \\*
MCG~--6-30-15 & 72013000 &           23 Jul 94 & 1 & 161.
  & $6.42^{+0.05}_{-0.05}$ & $\phantom{1}3.9^{+0.6}_{-0.6}$
  & $\phantom{1}78^{+ 11}_{- 11}$ &           110 & $50.3^{+0.6}_{-0.5}$ \\*
MCG~--6-30-15 & 75006000 & \phantom{1}3 Aug 97 & 1 & 85.1
  & $6.42^{+0.07}_{-0.07}$ & $\phantom{1}3.2^{+0.7}_{-0.7}$
  & $\phantom{1}83^{+ 20}_{- 19}$ & \phantom{1}45 & $39.9^{+0.7}_{-0.6}$ \\*
MCG~--6-30-15 & 75006010 & \phantom{1}7 Aug 97 & 1 & 85.8
  & $6.34^{+0.05}_{-0.05}$ & $\phantom{1}4.6^{+0.7}_{-0.7}$
  & $          121^{+ 19}_{- 19}$ & \phantom{1}94 & $37.3^{+0.6}_{-0.6}$ \\
IC~4329a      & 70005000 &           15 Aug 93 & 4 & 32.1
  & $6.37^{+0.05}_{-0.05}$ & $\phantom{1}9.3^{+2.1}_{-2.1}$
  & $\phantom{1}64^{+ 14}_{- 14}$ & \phantom{1}44 & $117.^{+2. }_{-2. }$ \\*
IC~4329a      & 75047000 & \phantom{1}7 Aug 97 & 1 & 20.1
  & $6.40^{+0.08}_{-0.08}$ & $\phantom{1}8.2^{+2.6}_{-2.6}$
  & $\phantom{1}62^{+ 20}_{- 20}$ & \phantom{1}21 & $112.^{+2. }_{-2. }$ \\*
IC~4329a      & 75047010 &           10 Aug 97 & 1 & 19.8
  & $6.36^{+0.08}_{-0.08}$ & $\phantom{1}8.8^{+2.4}_{-2.4}$
  & $\phantom{1}85^{+ 23}_{- 23}$ & \phantom{1}30 & $84.9^{+2.0}_{-2.0}$ \\*
IC~4329a      & 75047020 &           12 Aug 97 & 1 & 17.9
  & $6.40^{+0.12}_{-0.13}$ & $\phantom{1}8.0^{+2.6}_{-2.6}$
  & $\phantom{1}65^{+ 21}_{- 21}$ & \phantom{1}20 & $99.7^{+2.3}_{-2.3}$ \\*
IC~4329a      & 75047030 &           15 Aug 97 & 1 & 18.6
  & $6.44^{+0.10}_{-0.11}$ & $\phantom{1}7.8^{+2.8}_{-2.8}$
  & $\phantom{1}58^{+ 21}_{- 20}$ & \phantom{1}17 & $115.^{+2.4}_{-2.3}$ \\
%Mrk~279   & 72028000 & 5 Dec 94  & 2 & $6.483^{+0.060}_{-0.068}$ & & \\
NGC~5548      & 70018000 &           27 Jul 93 & 4 & 28.5
  & $6.39^{+0.06}_{-0.07}$ & $\phantom{1}5.4^{+1.4}_{-1.4}$
  & $\phantom{1}77^{+ 20}_{- 19}$ & \phantom{1}34 & $59.8^{+1.1}_{-1.4}$ \\*
NGC~5548      & 74038000 &           27 Jun 96 & 1 & 18.6
  & $6.42^{+0.12}_{-0.13}$ & $\phantom{1}5.9^{+2.2}_{-2.2}$
  & $\phantom{1}71^{+ 26}_{- 26}$ & \phantom{1}16 & $75.5^{+2.0}_{-1.9}$ \\*
NGC~5548      & 74038010 &           29 Jun 96 & 1 & 15.4
  & $6.39$\tablenotemark{c}\phantom{$_{-0.00}$}
                           & $\phantom{1}1.6^{+2.1}_{-1.6}$
  & $\phantom{1}20^{+ 27}_{- 20}$ & \phantom{11}2 & $66.4^{+2.2}_{-1.8}$ \\*
NGC~5548      & 74038020 & \phantom{1}1 Jul 96 & 1 & 17.7
  & $6.40^{+0.09}_{-0.09}$ & $\phantom{1}5.6^{+2.0}_{-2.0}$
  & $\phantom{1}81^{+ 29}_{- 29}$ & \phantom{1}18 & $57.3^{+1.8}_{-1.7}$ \\*
NGC~5548      & 74038030 & \phantom{1}3 Jul 96 & 1 & 17.6
  & $6.39$\tablenotemark{c}\phantom{$_{-0.00}$}
                           & $\phantom{1}1.7^{+1.8}_{-1.7}$
  & $\phantom{1}23^{+ 24}_{- 23}$ & \phantom{11}4 & $52.1^{+1.6}_{-1.9}$ \\*
NGC~5548      & 74038040 & \phantom{1}4 Jul 96 & 1 & 20.5
  & $6.39$\tablenotemark{c}\phantom{$_{-0.00}$}
                           & $\phantom{1}1.7^{+1.8}_{-1.7}$
  & $\phantom{1}23^{+ 25}_{- 23}$ & \phantom{11}4 & $60.5^{+1.6}_{-1.9}$ \\*
NGC~5548      & 76029000 &           15 Jun 98 & 1 & 21.7
  & $6.39$\tablenotemark{c}\phantom{$_{-0.00}$}
                           & $\phantom{1}0.6^{+1.8}_{-0.6}$
  & $          7.4^{+20.}_{-9.0}$ & \phantom{11}1 & $70.1^{+1.6}_{-1.6}$ \\*
NGC~5548      & 76029010 &           20 Jun 98 & 1 & 99.0
  & $6.34^{+0.06}_{-0.05}$ & $\phantom{1}3.9^{+0.9}_{-0.9}$
  & $\phantom{1}45^{+ 10}_{- 10}$ & \phantom{1}40 & $82.0^{+0.8}_{-0.8}$ \\
Mrk~841       & 70009000 &           22 Aug 93 & 2 & 31.0
  & $6.43^{+0.07}_{-0.06}$ & $\phantom{1}2.7^{+0.8}_{-0.8}$
  & $          170^{+ 52}_{- 52}$ & \phantom{1}26 & $14.0^{+0.8}_{-0.7}$ \\*
Mrk~841       & 71040000 &           21 Feb 94 & 2 & 20.8
  & $6.41^{+0.15}_{-0.12}$ & $\phantom{1}1.8^{+0.9}_{-0.9}$
  & $          125^{+ 63}_{- 62}$ & \phantom{1}10 & $12.4^{+0.9}_{-0.8}$ \\*
Mrk~841       & 75010000 &           31 Jul 97 & 1 & 39.4
  & $6.42$\tablenotemark{c}\phantom{$_{-0.00}$}
                           & $\phantom{1}1.1^{+0.8}_{-0.8}$
  & $\phantom{1}46^{+ 32}_{- 33}$ & \phantom{11}5 & $19.7^{+0.6}_{-0.7}$ \\
Mrk~509       & 71013000 &           29 Apr 94 & 1 & 41.4
  & $6.42^{+0.12}_{-0.14}$ & $\phantom{1}2.7^{+1.2}_{-1.2}$
  & $\phantom{1}43^{+ 18}_{- 18}$ & \phantom{1}12 & $51.0^{+1.4}_{-0.9}$ \\*
Mrk~509       & 74024000 &           20 Oct 96 & 1 & 7.12
  & $6.48$\tablenotemark{c}\phantom{$_{-0.00}$}
                           & $\phantom{1}0.0^{+2.0}_{-0  }$
  & $          0.0^{+ 22}_{-0. }$ & \phantom{11}3 & $63.7^{+2.8}_{-3.1}$ \\*
Mrk~509       & 74024010 &           22 Oct 96 & 1 & 2.47
  & $6.48$\tablenotemark{c}\phantom{$_{-0.00}$}
                           & $\phantom{1}3.4^{+6.2}_{-3.4}$
  & $\phantom{1}40^{+ 72}_{- 40}$ & \phantom{11}1 & $55.6^{+5.3}_{-4.6}$ \\*
Mrk~509       & 74024020 &           25 Oct 96 & 1 & 8.42
  & $6.28^{+0.11}_{-0.10}$ & $\phantom{1}6.7^{+2.9}_{-2.9}$
  & $\phantom{1}57^{+ 25}_{- 25}$ & \phantom{1}11 & $66.4^{+2.7}_{-2.6}$ \\*
Mrk~509       & 74024030 &           28 Oct 96 & 1 & 11.7
  & $6.48$\tablenotemark{c}\phantom{$_{-0.00}$}
                           & $\phantom{1}4.0^{+2.8}_{-2.8}$
  & $\phantom{1}38^{+ 27}_{- 27}$ & \phantom{11}4 & $78.1^{+2.3}_{-2.8}$ \\*
Mrk~509       & 74024040 &           31 Oct 96 & 1 & 11.0
  & $6.48$\tablenotemark{c}\phantom{$_{-0.00}$}
                           & $\phantom{1}3.4^{+2.4}_{-2.4}$
  & $\phantom{1}51^{+ 37}_{- 37}$ & \phantom{11}5 & $57.8^{+2.0}_{-2.4}$ \\*
Mrk~509       & 74024050 & \phantom{1}4 Nov 96 & 1 & 12.3
  & $6.47^{+0.08}_{-0.09}$ & $\phantom{1}6.3^{+2.3}_{-2.3}$
  & $          113^{+ 41}_{- 41}$ & \phantom{1}15 & $49.3^{+2.0}_{-1.9}$ \\*
Mrk~509       & 74024060 & \phantom{1}6 Nov 96 & 1 & 10.6
  & $6.48$\tablenotemark{c}\phantom{$_{-0.00}$}
                           & $\phantom{1}0.3^{+2.1}_{-0.3}$
  & $          5.9^{+39.}_{-5.9}$ & \phantom{11}4 & $43.4^{+1.9}_{-2.1}$ \\*
Mrk~509       & 74024070 & \phantom{1}9 Nov 96 & 1 & 9.74
  & $6.26^{+0.32}_{-0.09}$ & $\phantom{1}5.4^{+2.5}_{-2.5}$
  & $\phantom{1}93^{+ 43}_{- 43}$ & \phantom{1}11 & $49.9^{+2.2}_{-2.1}$ \\*
Mrk~509       & 74024080 &           12 Nov 96 & 1 & 10.1
  & $6.96^{+0.21}_{-0.15}$ & $\phantom{1}5.1^{+2.8}_{-2.8}$
  & $\phantom{1}88^{+ 49}_{- 49}$ & \phantom{11}8 & $57.4^{+2.3}_{-2.2}$ \\*
Mrk~509       & 74024090 &           16 Nov 96 & 1 & 9.19
  & $6.48$\tablenotemark{c}\phantom{$_{-0.00}$}
                           & $\phantom{1}3.5^{+2.5}_{-2.5}$
  & $\phantom{1}52^{+ 37}_{- 36}$ & \phantom{11}4 & $46.7^{+2.2}_{-2.0}$ \\
NGC~7469      & 71028000 &           24 Nov 93 & 2 & 9.15
  & $6.61^{+0.13}_{-0.16}$ & $\phantom{1}3.8^{+2.0}_{-1.9}$
  & $\phantom{1}91^{+ 47}_{- 46}$ & \phantom{11}8 & $35.8^{+1.8}_{-1.7}$ \\*
NGC~7469      & 71028030 &           26 Nov 93 & 2 & 10.5
  & $6.44^{+0.06}_{-0.07}$ & $\phantom{1}4.0^{+1.8}_{-1.8}$
  & $          113^{+ 49}_{- 49}$ & \phantom{1}11 & $27.9^{+1.6}_{-1.4}$ \\*
NGC~7469      & 71028010 & \phantom{1}2 Dec 93 & 2 & 19.0
  & $6.38^{+0.07}_{-0.08}$ & $\phantom{1}2.1^{+1.3}_{-1.2}$
  & $\phantom{1}47^{+ 28}_{- 28}$ & \phantom{11}6 & $35.7^{+1.0}_{-1.3}$ \\*
NGC~7469      & 15030000 &           26 Jun 94 & 1 & 11.7
  & $6.27^{+0.09}_{-0.09}$ & $\phantom{1}4.6^{+1.8}_{-1.8}$
  & $\phantom{1}80^{+ 32}_{- 32}$ & \phantom{1}14 & $38.0^{+1.6}_{-1.5}$ \\
MCG~--2-58-22 & 70004000 &           25 May 93 & 4 & 18.3
  & $6.40^{+0.22}_{-0.11}$ & $\phantom{1}2.6^{+1.2}_{-1.2}$
  & $\phantom{1}88^{+ 40}_{- 40}$ & \phantom{1}10 & $18.4^{+0.9}_{-1.0}$ \\*
MCG~--2-58-22 & 75049000 & \phantom{1}1 Jun 97 & 1 & 37.9
  & $6.33$\tablenotemark{c}\phantom{$_{-0.00}$}
                           & $\phantom{1}.05^{+1.1}_{-.05}$
  & $          0.9^{+20.}_{-0.9}$ & \phantom{11}2 & $36.7^{+1.1}_{-0.9}$ \\*
MCG~--2-58-22 & 75049010 &           15 Dec 97 & 1 & 38.3
  & $6.27^{+0.09}_{-0.09}$ & $\phantom{1}2.8^{+1.0}_{-1.0}$
  & $\phantom{1}66^{+ 24}_{- 24}$ & \phantom{1}17 & $34.5^{+0.9}_{-0.9}$ \\
%Mrk~279   & 72028000 & 5 Dec 94  & 2 & $6.483^{+0.060}_{-0.068}$ & & \\
\enddata
\tablenotetext{a}{The extremely low EW reported for this data set is an 
    artifact of a particularly large absorption correction.  When EW is
    measured without an absorption correction, the value obtained is
    $306\pm84$~eV.  See \S3.3.2.}
\tablenotetext{b}{NGC~4151 fits utilized 2--10~keV data in order to
    reduce the impact of the warm absorber, which complicates the
    interpretation of the low energy end of the spectrum.}
\tablenotetext{c}{When the Fe line was undetected ($\Delta\chi^2 < 6$ 
    after removing the narrow Gaussian from the model),
    we fixed the line energy at the mean value in the significant
    detections and found upper limits on the line flux and EW.}
\tablecomments{Fe~K line parameters from narrow Gaussian fits.  Our
    fits were performed on {\it ASCA} data in the 1--10~keV range,
    using an absorbed double power law model with a narrow Gaussian
    component.  The errors represent $90\%$ confidence limits
    ($\Delta\chi^2 = 2.706$ for one interesting parameter).  The
    columns are (1)~galaxy name, (2)~{\it ASCA} sequence number,
    (3)~date of the start of each observation, (4)~CCD mode of SIS
    detectors in the utilized data, (5)~SIS-0 integration time,
    (6)~(rest frame) energy of the Gaussian fit to the Fe~K$\alpha$
    feature, (7)~Gaussian normalization, (8)~Fe equivalent width,
    (9)~the change in $\chi^{2}$ when the Gaussian component is
    excluded from the model, and (10)~2--10~keV continuum flux
    ($\times 10^{-12}$~\cgsflux).}
\end{deluxetable}

\clearpage

\begin{deluxetable}{lccccccccc}
\tablecolumns{10}
\tablewidth{0pc}
\tablecaption{Summary of Iron Line Variability\label{tab:variabSummary}}
\tablehead{
              & \multicolumn{2}{c}{number} 
	& \multicolumn{4}{c}{Parameter significance} & & &  \\
Galaxy name   & obs & non 
	& E & \mathsub{F}{Fe} & EW & \mathsub{F}{2-10}
	& Log(\mathsub{L}{X}) & Log(\mathsub{F}{2-10}) 
	& \mathsub{\Delta t}{min} }
\startdata
Mkn~509       & 11 & 6 & 95\% & 95\% & 99\% & 99\% & 44.11 & --10.25 & 2.5~d  \\
MCG~--2-58-22 & 3  & 1 & ---  & 95\% & 95\% & 99\% & 44.10 & --10.52 & 0.5~yr \\
Fairall~9     & 8  & 0 & 99\% & 68\% & 68\% & 99\% & 44.02 & --10.61 & 4~d    \\
IC~4329a      & 5  & 0 & ---  & ---  & 68\% & 99\% & 43.72 &  --9.98 & 2~d    \\
{\it Mkn~279} & {\it 1}
                   & {\it 0}
                       & {\it 95\%}
                              & {\it 68\%}
                                     & {\it 90\%}
                                            & {\it 99\%}
                                                   & {\it 43.72}
                                                           & {\it --10.53}
                                                                     & {\it 10~ks} \\
Mkn~841       & 3  & 1 & ---  & 68\% & 95\% & 99\% & 43.59 & --10.81 & 0.5~yr \\
NGC~5548      & 8  & 4 & ---  & 95\% & 99\% & 99\% & 43.55 & --10.18 & 1.5~d  \\
MCG~8-11-11   & 2  & 0 & ---  & ---  & ---  & ---  & 43.30 & --10.58 & 2.5~d  \\
NGC~7469      & 4  & 0 & 95\% & 68\% & 68\% & 99\% & 43.24 & --10.46 & 2~d    \\
NGC~3783      & 6  & 0 & 95\% & 68\% & 95\% & 99\% & 43.01 & --10.13 & 1.5~d  \\
NGC~3516      & 4  & 0 & 90\% & 95\% & 95\% & 99\% & 42.94 & --10.22 & 2~d    \\
NGC~4593      & 2  & 0 & ---  & ---  & ---  & 99\% & 42.85 & --10.34 & 3.5~yr \\
MCG~--6-30-15 & 5  & 0 & 90\% & 68\% & 95\% & 99\% & 42.70 & --10.35 & 4~d    \\
NGC~4151      & 7  & 0 & ---  & ---  & 95\% & 99\% & 42.68 &  --9.64 & 2~d    \\
NGC~3227      & 2  & 0 & ---  & ---  & ---  & 99\% & 41.97 & --10.47 & 2~yr   \\
NGC~4051      & 2  & 0 & ---  & ---  & ---  & 99\% & 41.47 & --10.57 & 1~yr   \\
\enddata
\tablecomments{Galaxies are listed by decreasing 2--10~keV luminosity.
    Note that Mkn~279 has been observed only once, and is therefore
    not included in our sample.  The columns are: (1)~galaxy name, 
    (2)~number of observations, (3)~number of non-detections of the
    Fe~K line, (4)~significance of line energy variability,
    (5)~significance of line flux variability, (6)~significance of
    equivalent width variability, (7)~significance of variability of
    2--10~keV flux, (8)~2--10~keV luminosity, in ergs~s$^{-1}$,
    (9)~2--10~keV flux, in \cgsflux , averaged over all observations,
    and (10)~shortest variability timescale probed for each galaxy.}
\end{deluxetable}

\clearpage

%\clearpage

\clearpage
 
\clearpage

%\clearpage

\clearpage

\addtocounter{figure}{1}

\begin{figure}
\epsscale{0.6}
\plotone{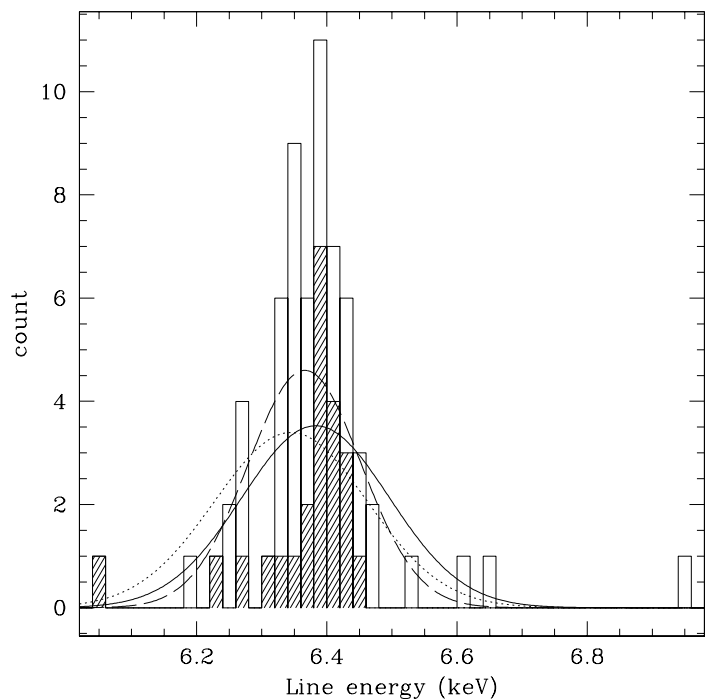}
\caption{Results of Gaussian fits to Fe~K$\alpha$ line for the
galaxies in our sample and the Nandra et~al.\ (1997) sample.  Shaded
bins correspond to our fits of the observations included in the N97
sample and the non-shaded bins correspond to galaxies and/or
observations not included in their sample.  Also drawn are the
best-fitting Gaussian distributions for our measurements of the
combined sample (solid curve; N=64, E=6.38~keV, $\sigma$=0.11~keV),
our fits to the N97 sample (dashed curve; N=23, E=6.37~keV,
$\sigma$=0.09~keV) and the N97 results (dotted curve; N=23,
E=6.34~keV, $\sigma$=0.12~keV).  Our result is similar to N97,
indicating that there is no model-dependent bias in our fitting
technique.
\label{fig:lineEdist}
}
\end{figure}

%\clearpage

\begin{figure}
\figurenum{8 a--d}
\epsscale{1.0}
\plottwo{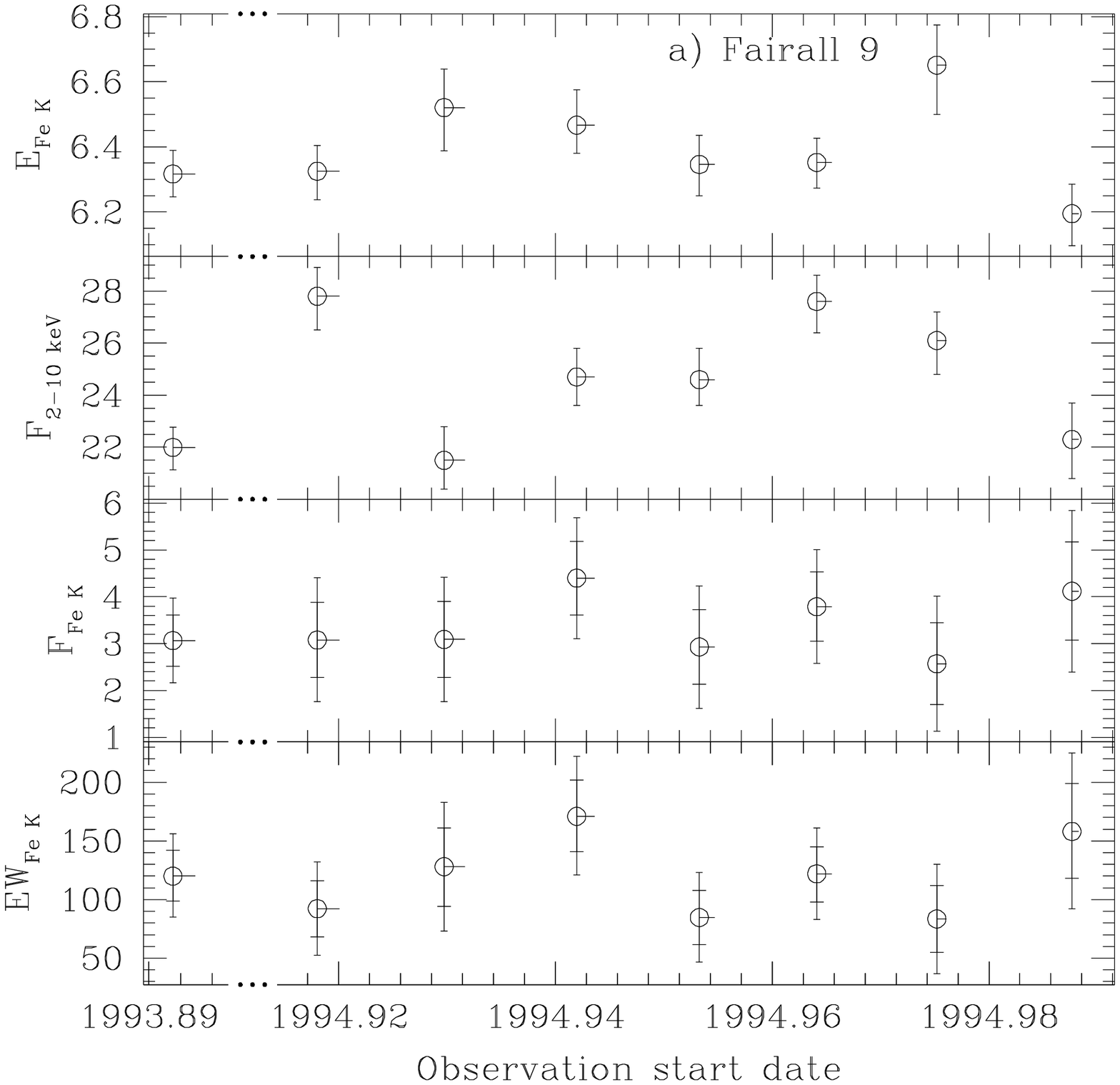}{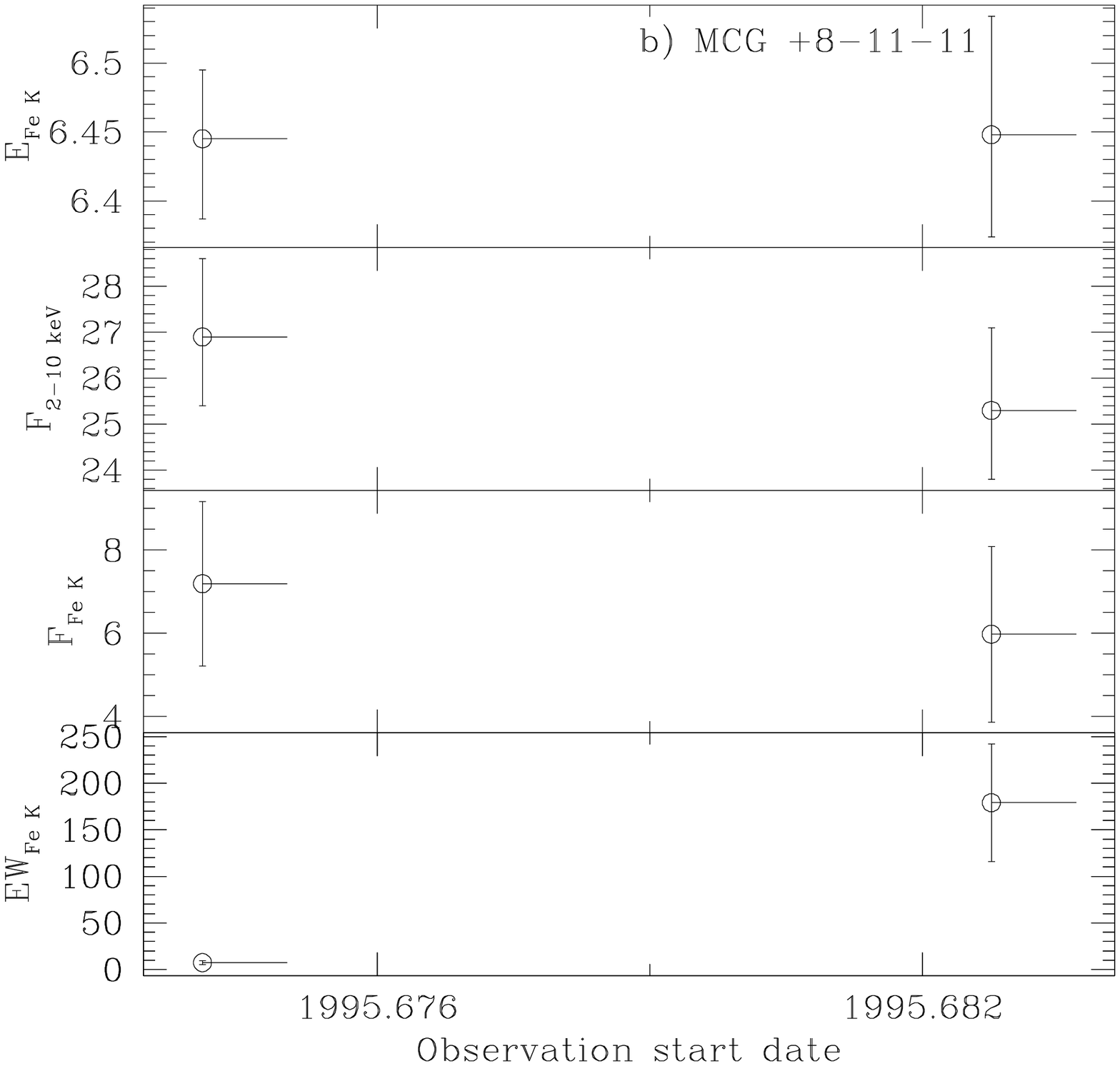}\\
\epsscale{2.2}
\plottwo{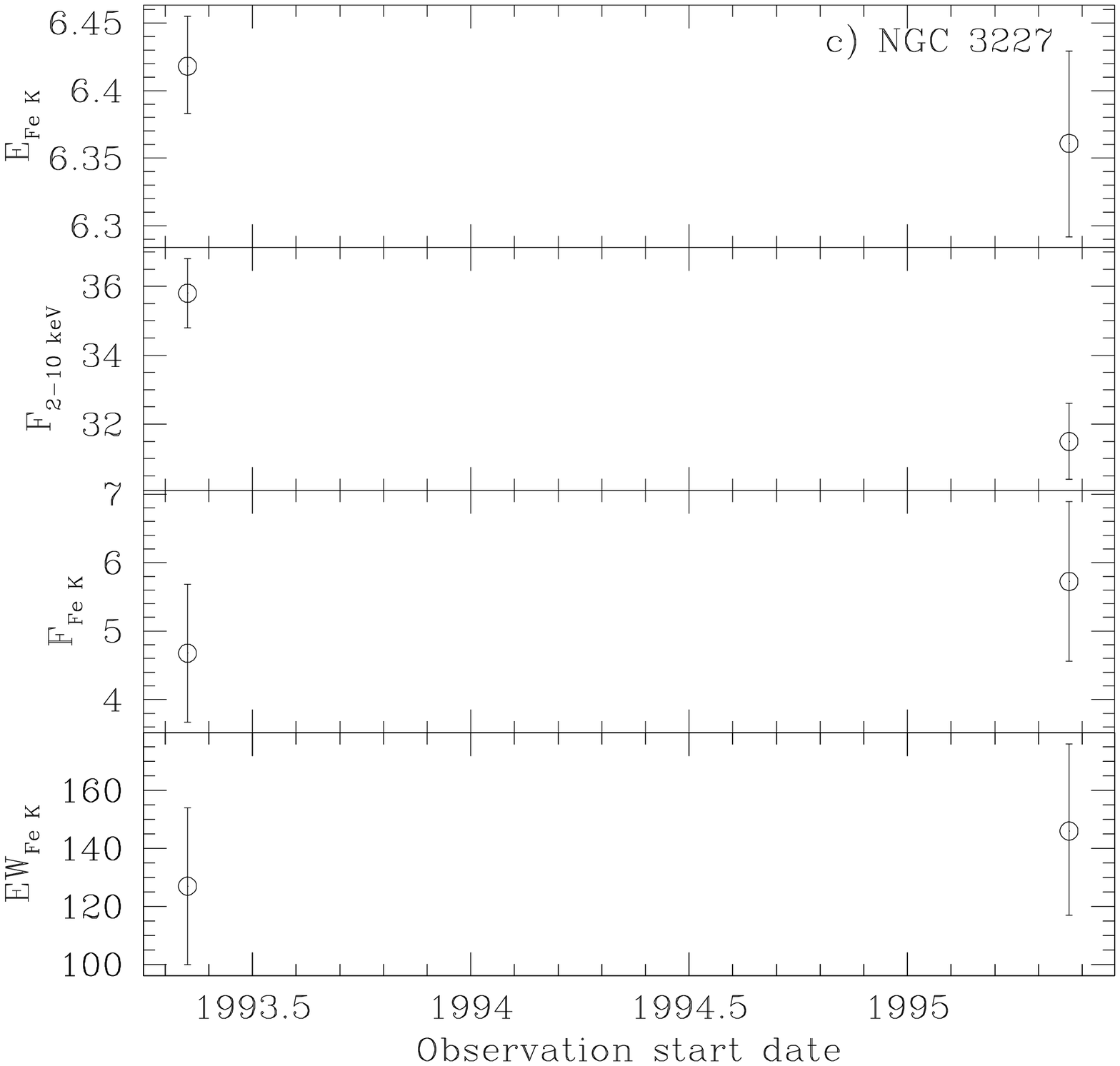}{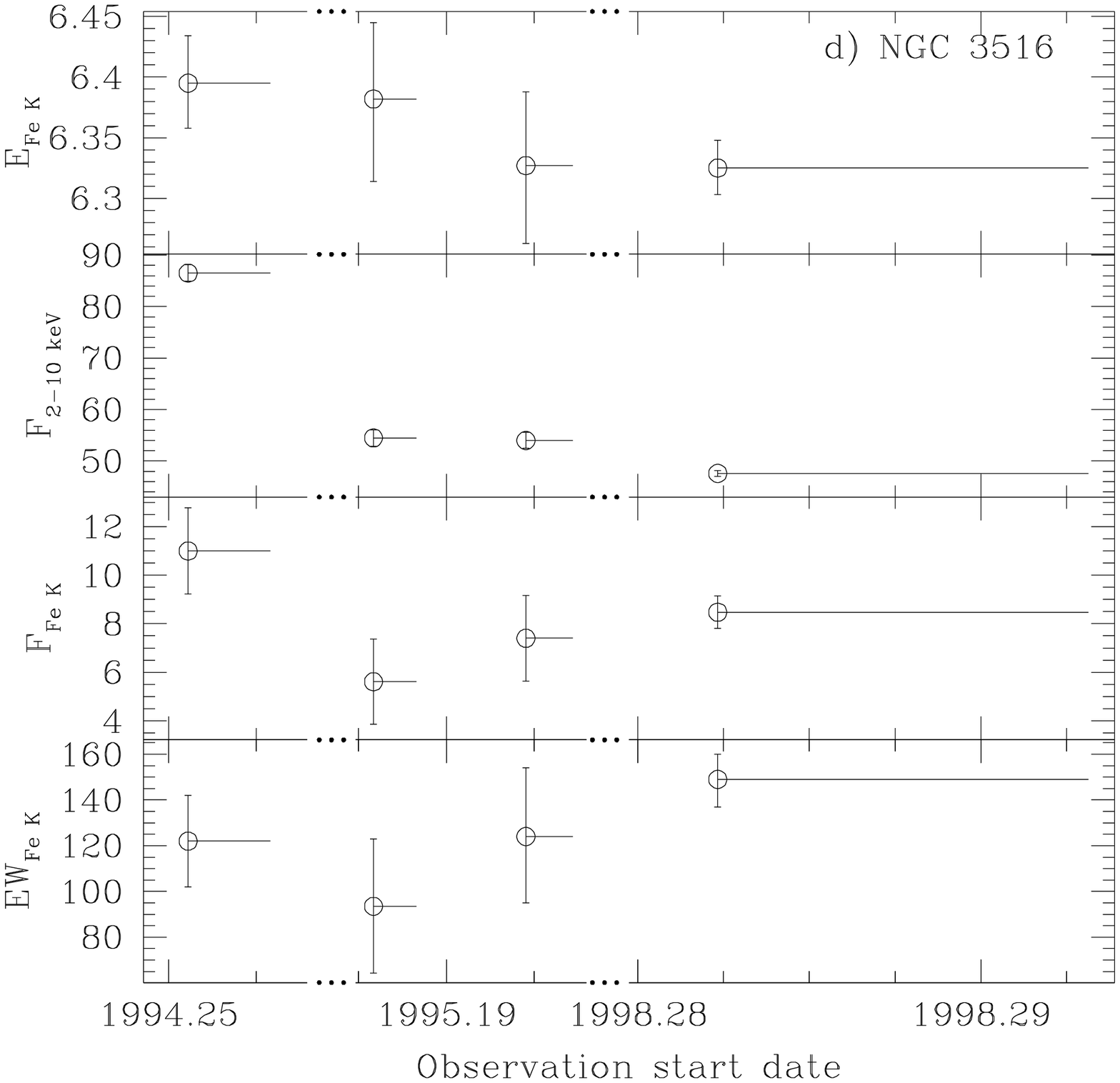}
\caption{Plotted from top to bottom in each panel are: Fe~K$\alpha$
line core energy (keV), 2--10~keV flux ($\times10^{-12}$~\cgsflux ),
line core flux (represented by the Gaussian model normalization,
$\times10^{-5}$) and line core EW (eV), all plotted against the
observation date.  Error bars represent $90\%$ confidence.  Horizontal
lines represent the observation duration.
\mbox{(a) Fairall 9:} $68\%$ errors are also shown for line flux and EW.  
Note that the time axis is broken, and small ticks =~1~day. 
\mbox{(b) MCG 8-11-11:} Note that the small EW of the first observation is
likely to be an artifact of our model; a more accurate estimate is
$306\pm84$~eV.  See \S3.3.2 of text.  Small ticks =~1~day.
\mbox{(c) NGC 3227.}
\mbox{(d) NGC 3516:} Note that the time axis is broken; small ticks =~1~day.
\label{fig:flux_var}
}
\end{figure}

%\clearpage 

\begin{figure}
\figurenum{8 e--h}
\epsscale{1.0}
\plottwo{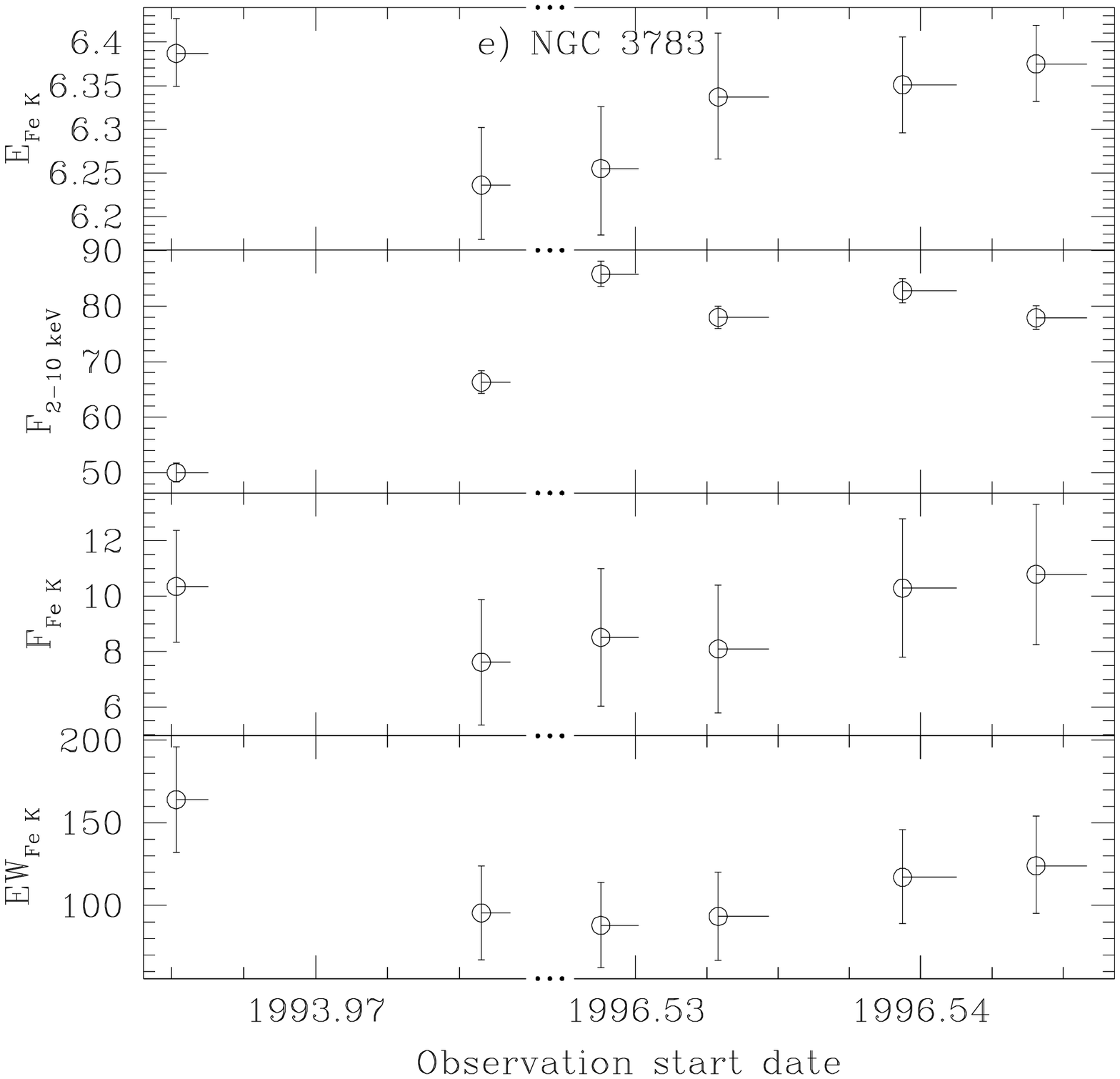}{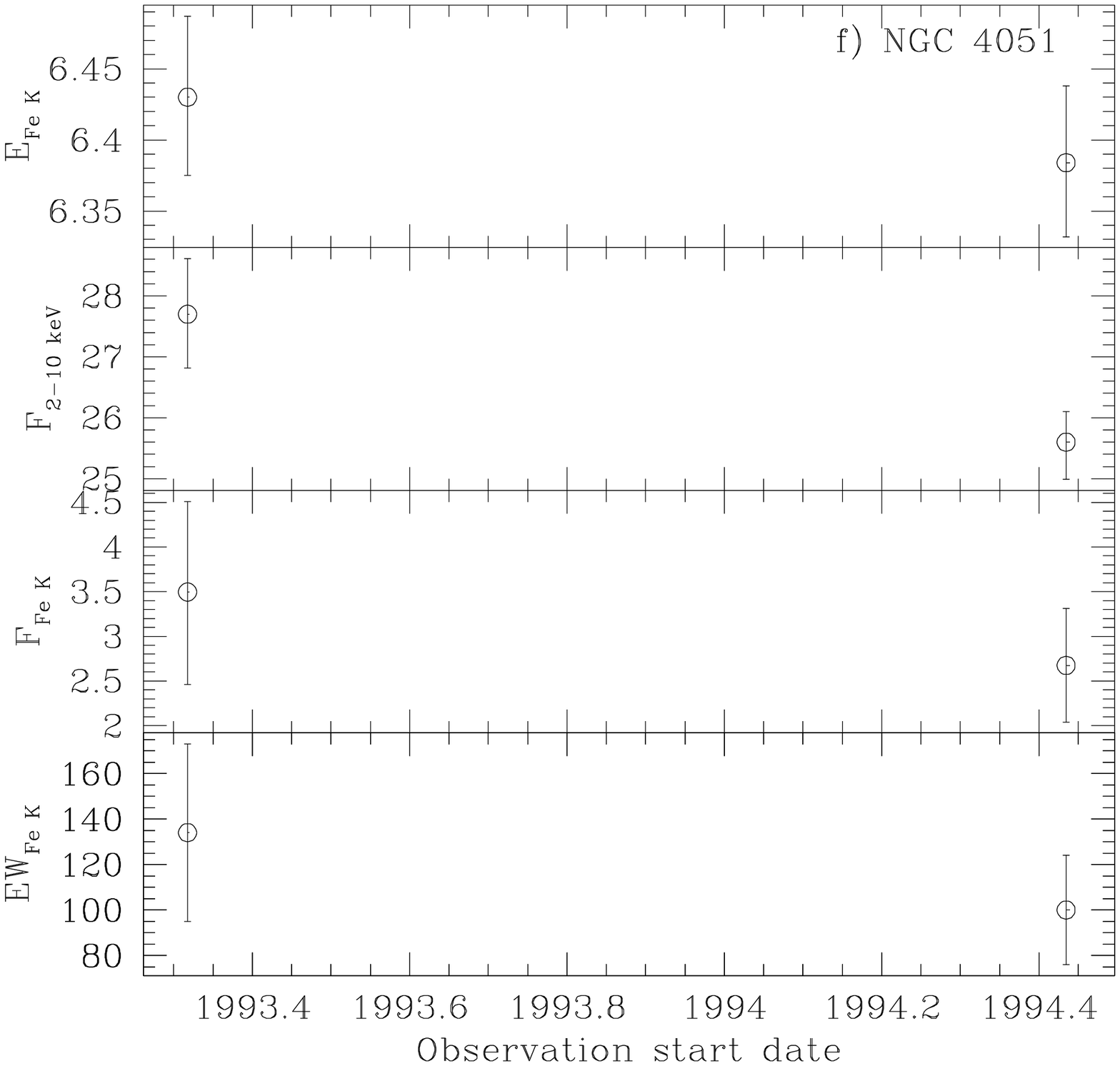} \\
\epsscale{2.2}
\plottwo{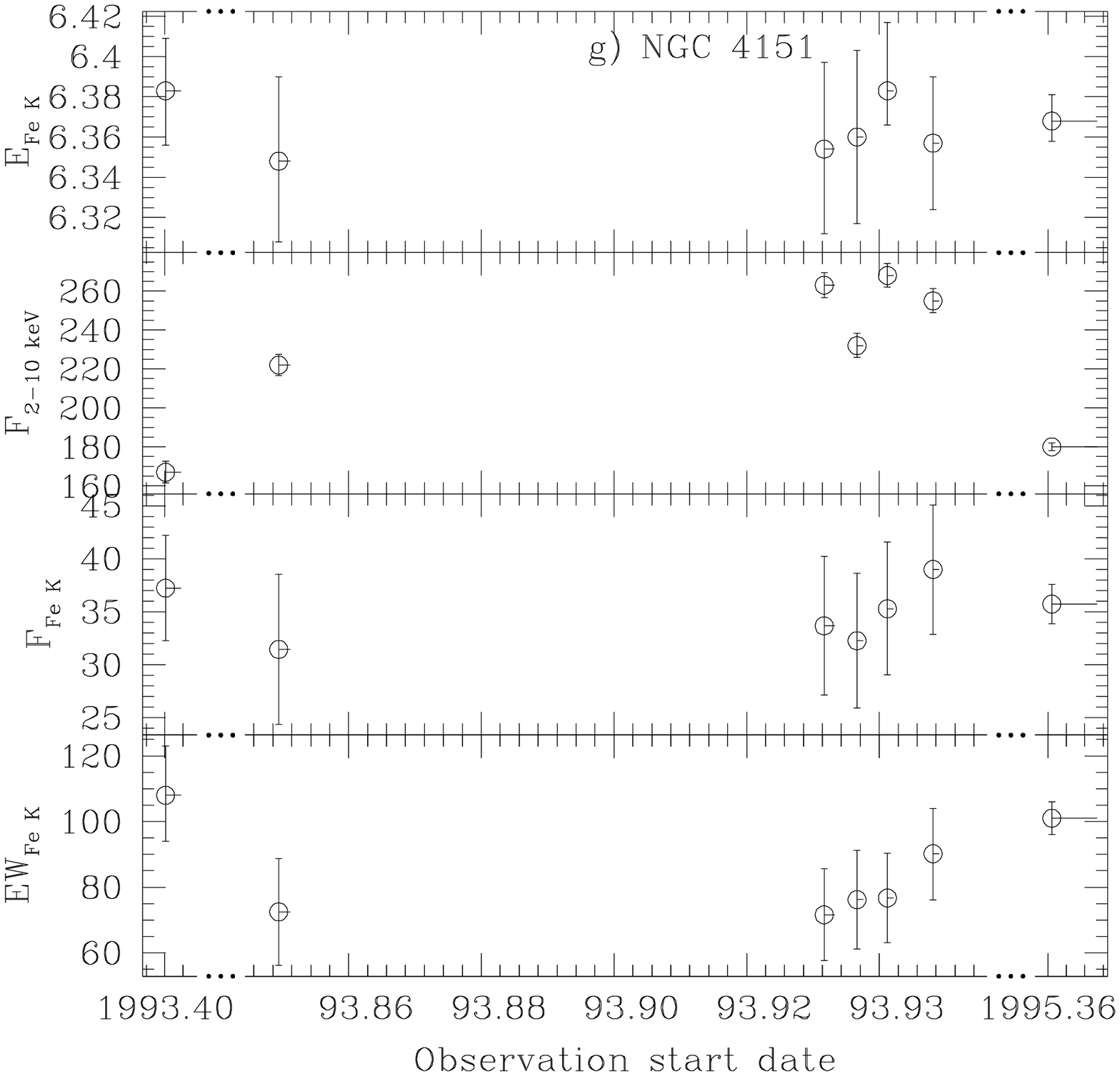}{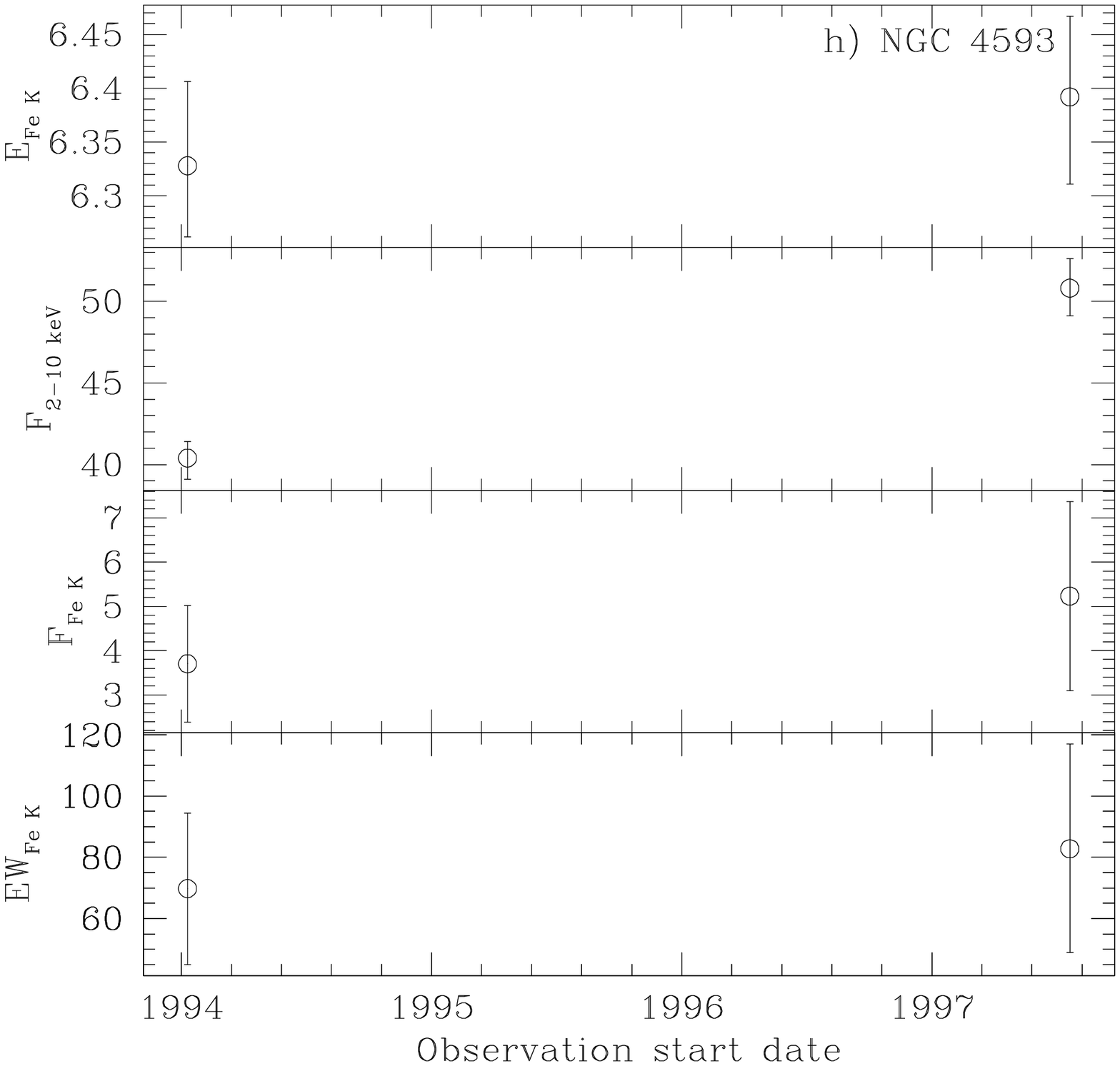}
\caption{
\mbox{(e) NGC 3783.}
\mbox{(f) NGC 4051.}
\mbox{(g) NGC 4151.}
\mbox{(h) NGC 4593.}
Note that for panels $e$ and $g$, the time axis is broken, and is
scaled so that the small ticks =~1~day.
}
\end{figure}

%\clearpage 

\begin{figure}
\figurenum{8 i--l}
\epsscale{1.0}
\plottwo{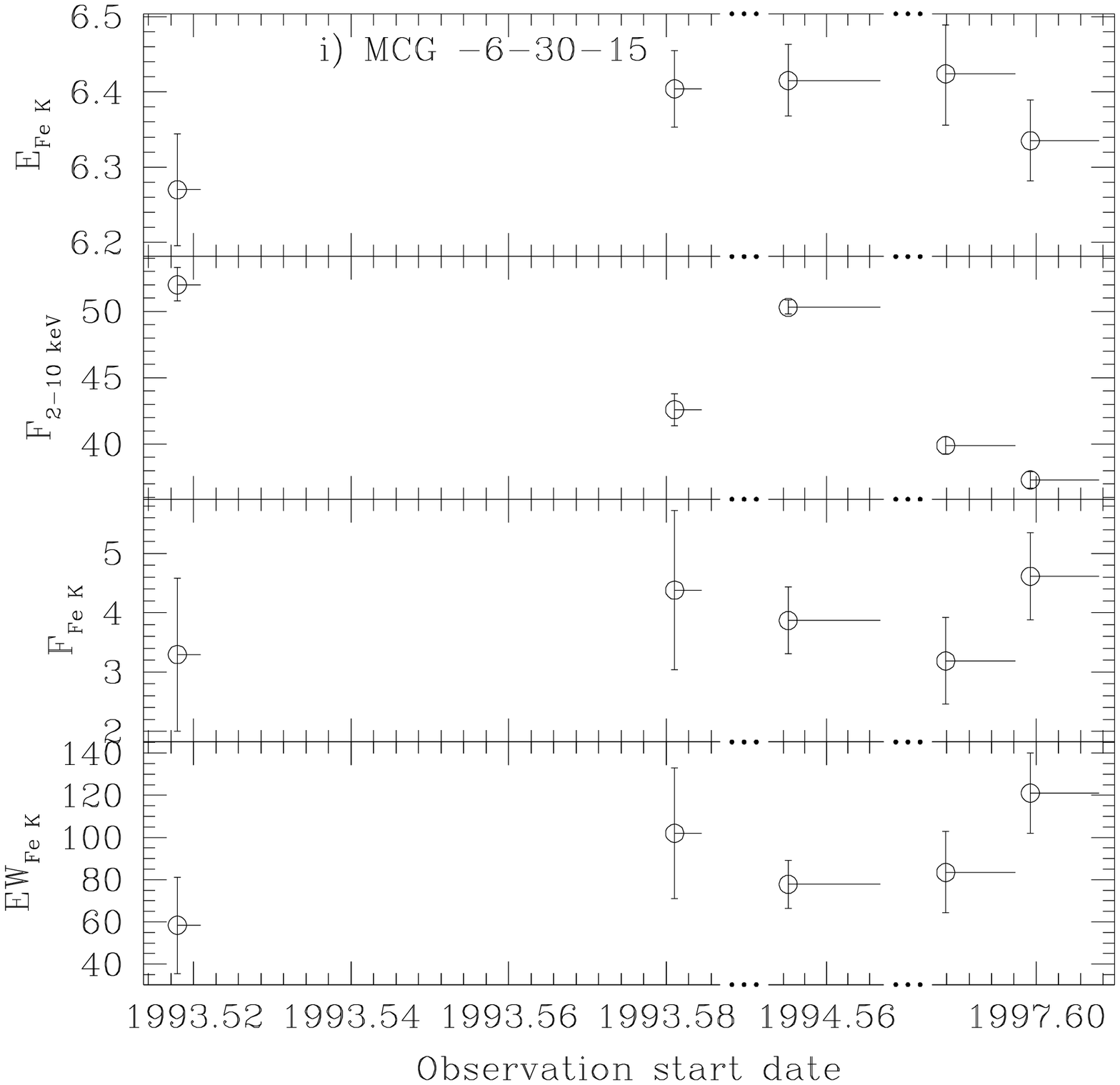}{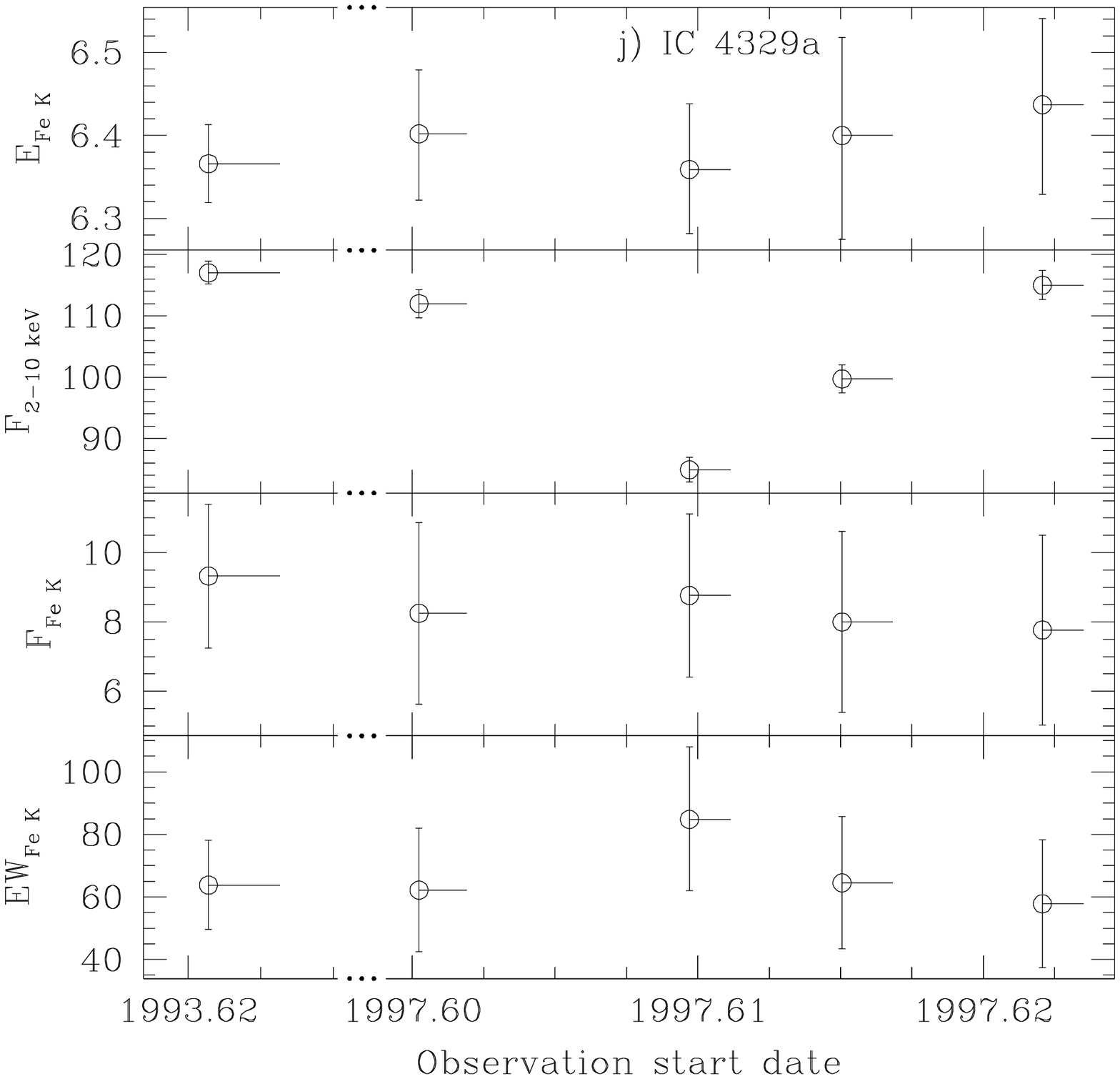} \\
\epsscale{2.2}
\plottwo{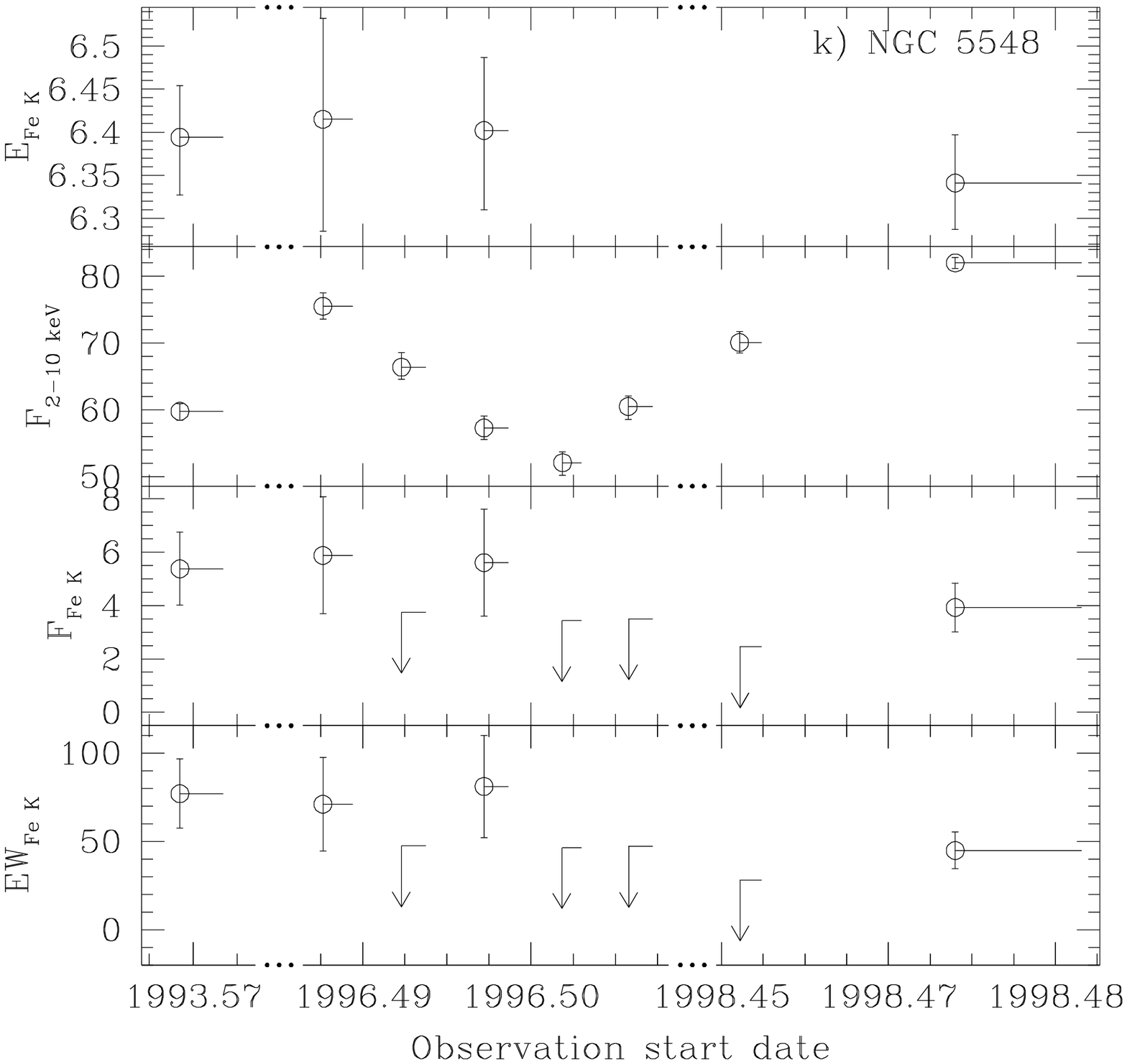}{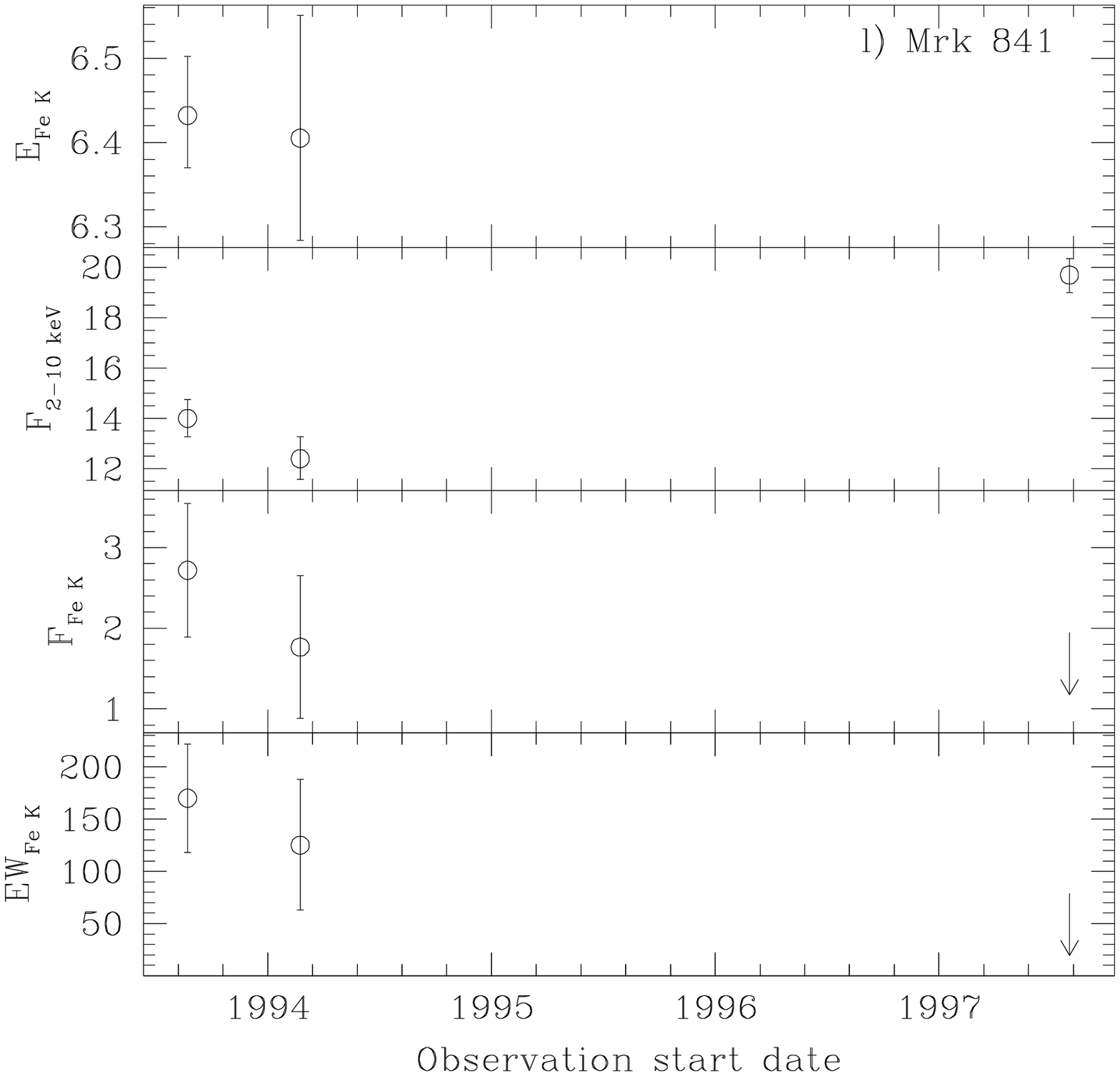}
\caption{
\mbox{(i) MCG --6-30-15.}
\mbox{(j) IC 4329a.}
\mbox{(k) NGC 5548.}
\mbox{(l) Mrk 841.}
Note that for panels $i-k$, the time axis is broken, and is
scaled so that the small ticks =~1~day.
}
\end{figure}

%\clearpage 

\begin{figure}
\figurenum{8 m--o}
\epsscale{1.0}
\plottwo{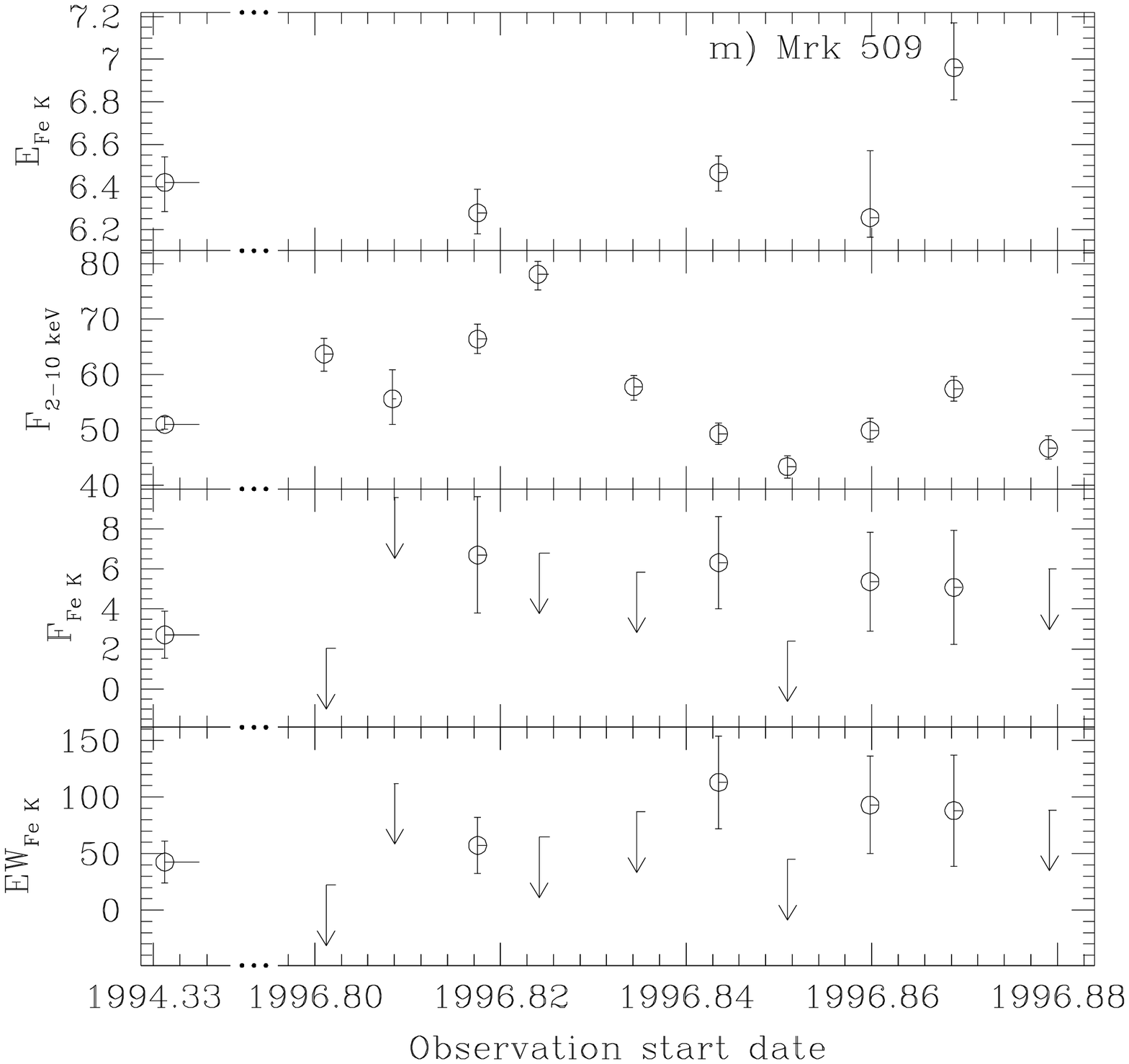}{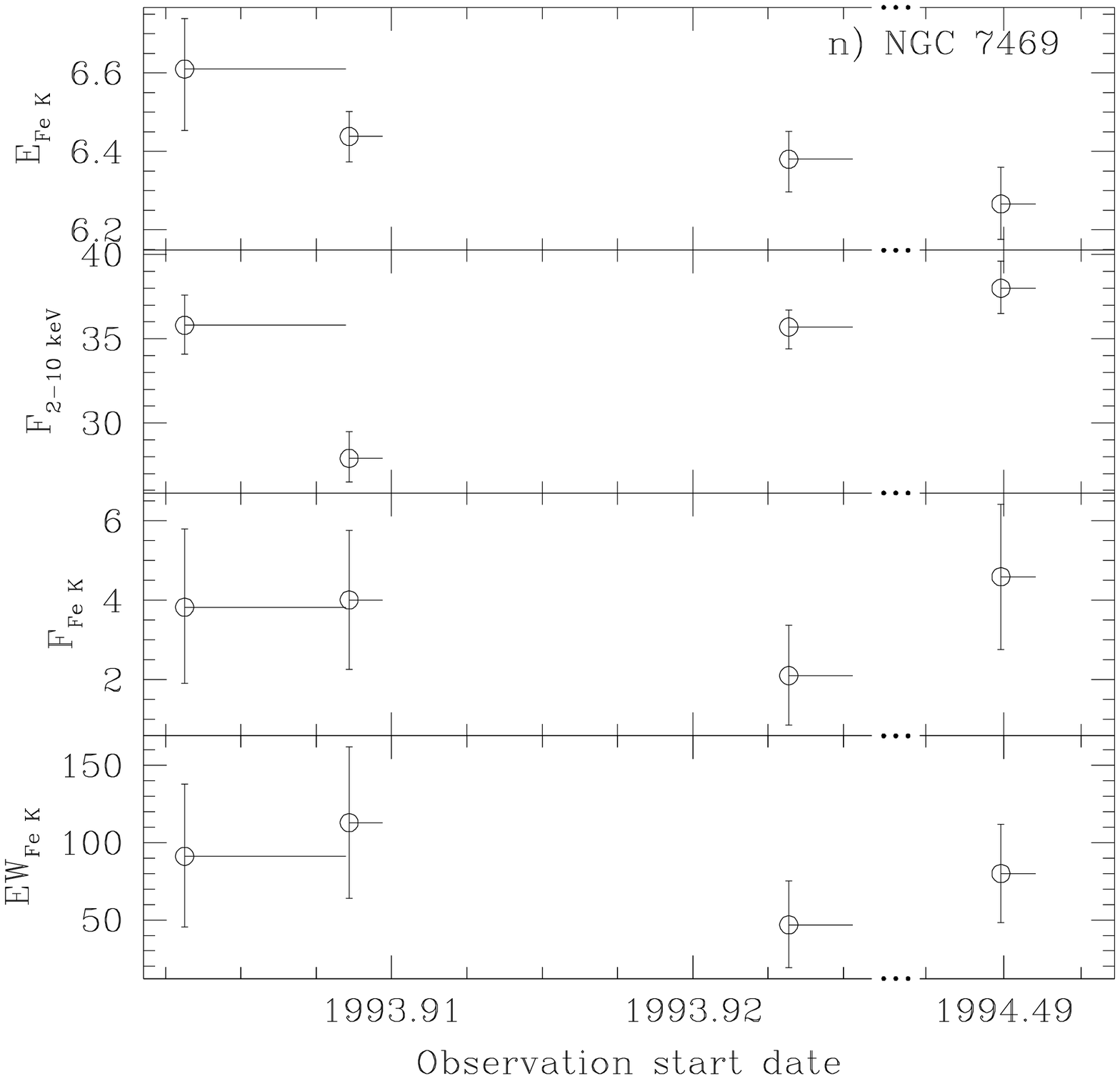} \\
\epsscale{1.0}
\plotone{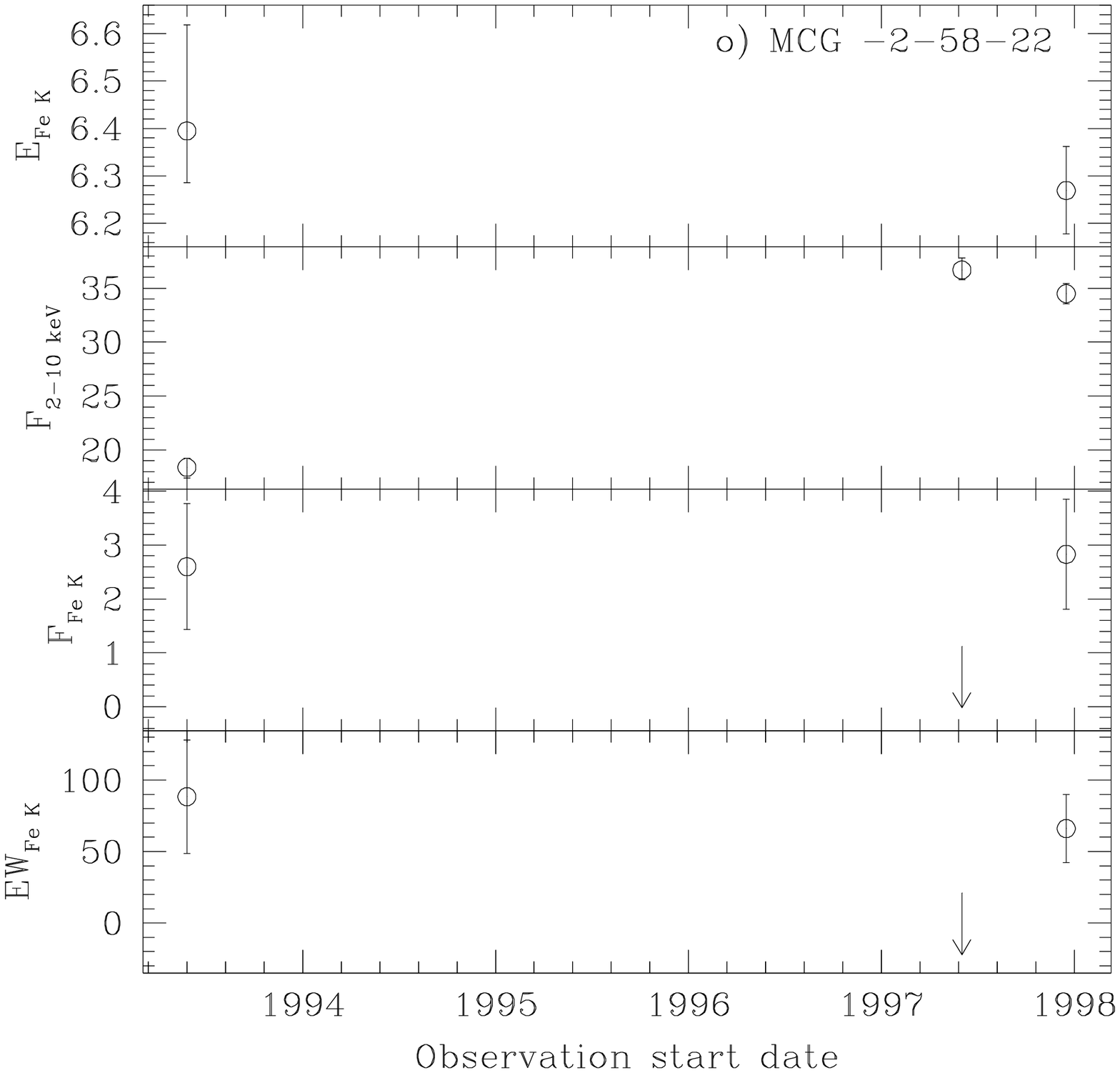}
\caption{
\mbox{(m) Mrk 509.}
\mbox{(n) NGC 7469.}
\mbox{(o) MCG --2-58-22.}
Note that for panels $m$ and $n$, the time axis is broken, and is
scaled so that the small ticks =~1~day.
}
\end{figure}

\addtocounter{figure}{1}

%\clearpage 

\begin{figure}
\figurenum{9 a--d}
\epsscale{1.0}
\plottwo{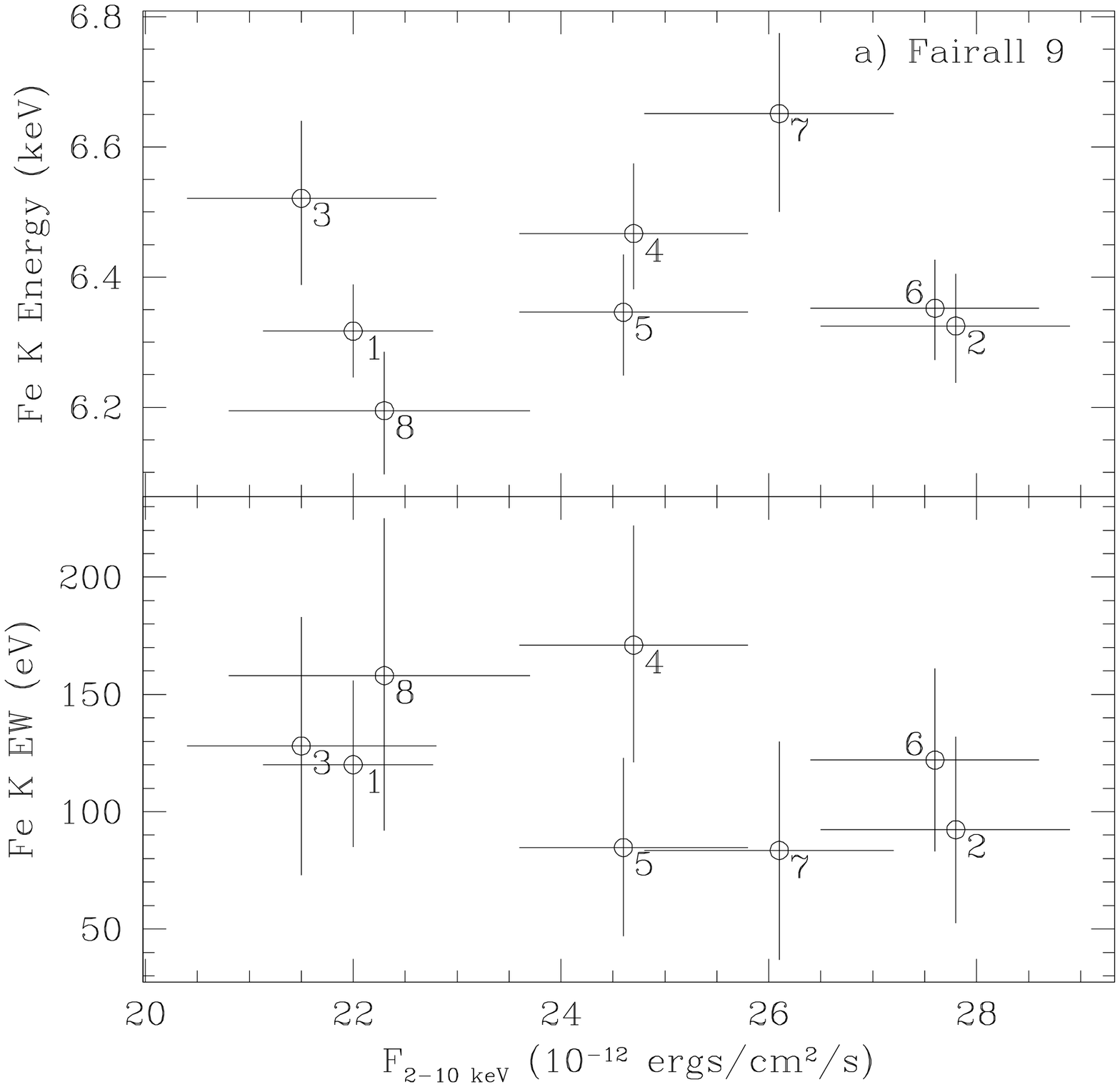}{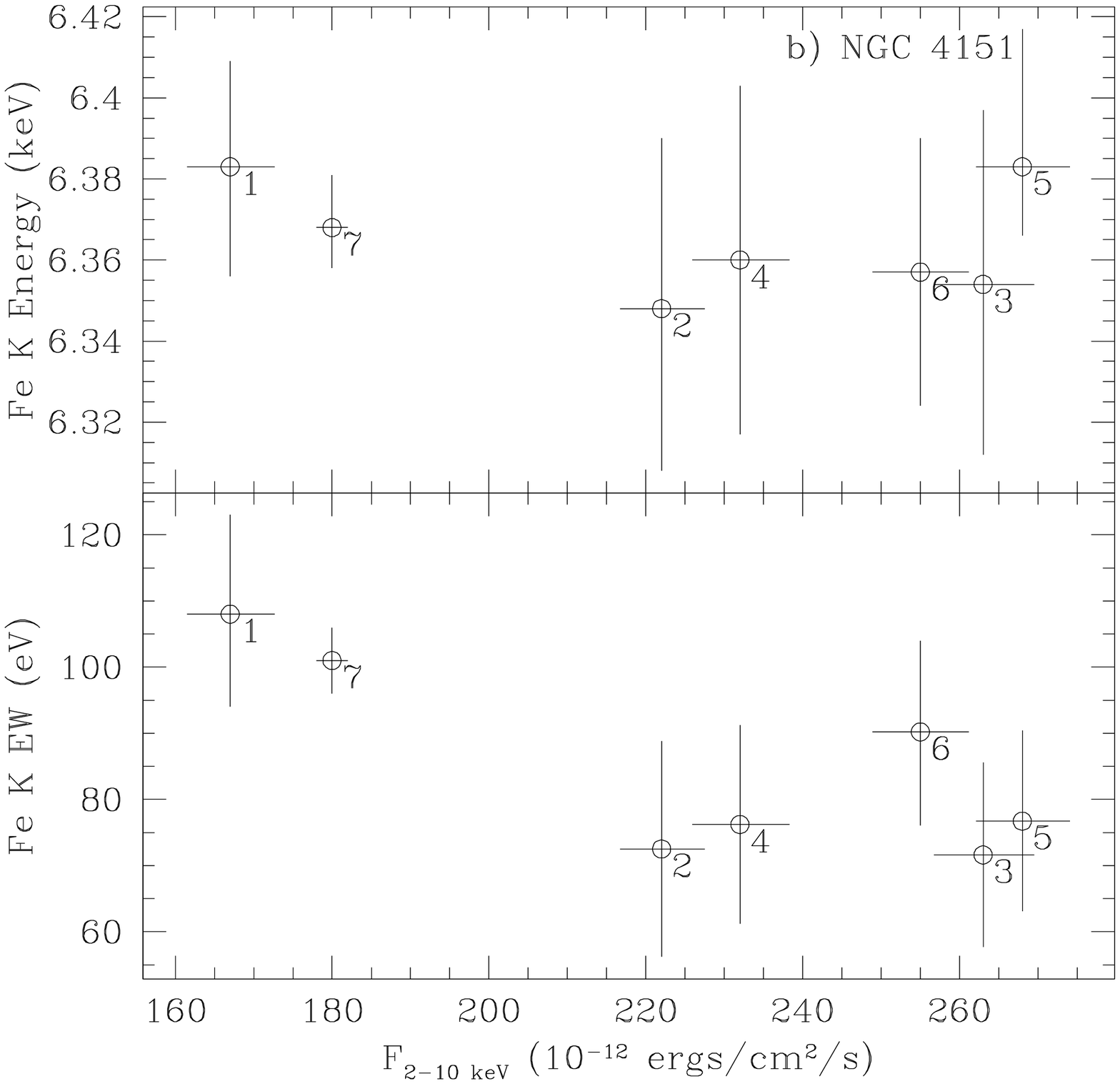} \\
\epsscale{2.2}
\plottwo{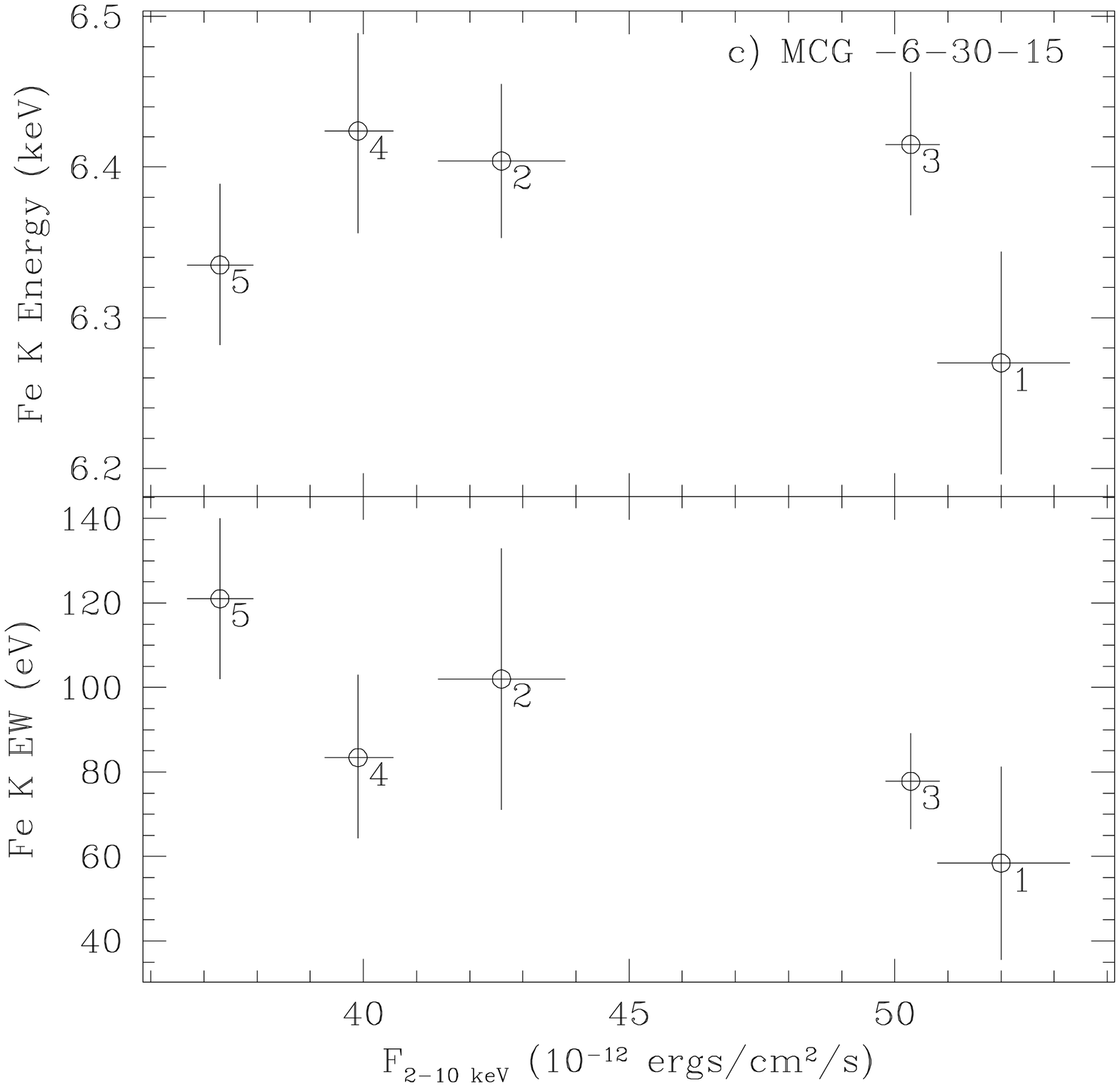}{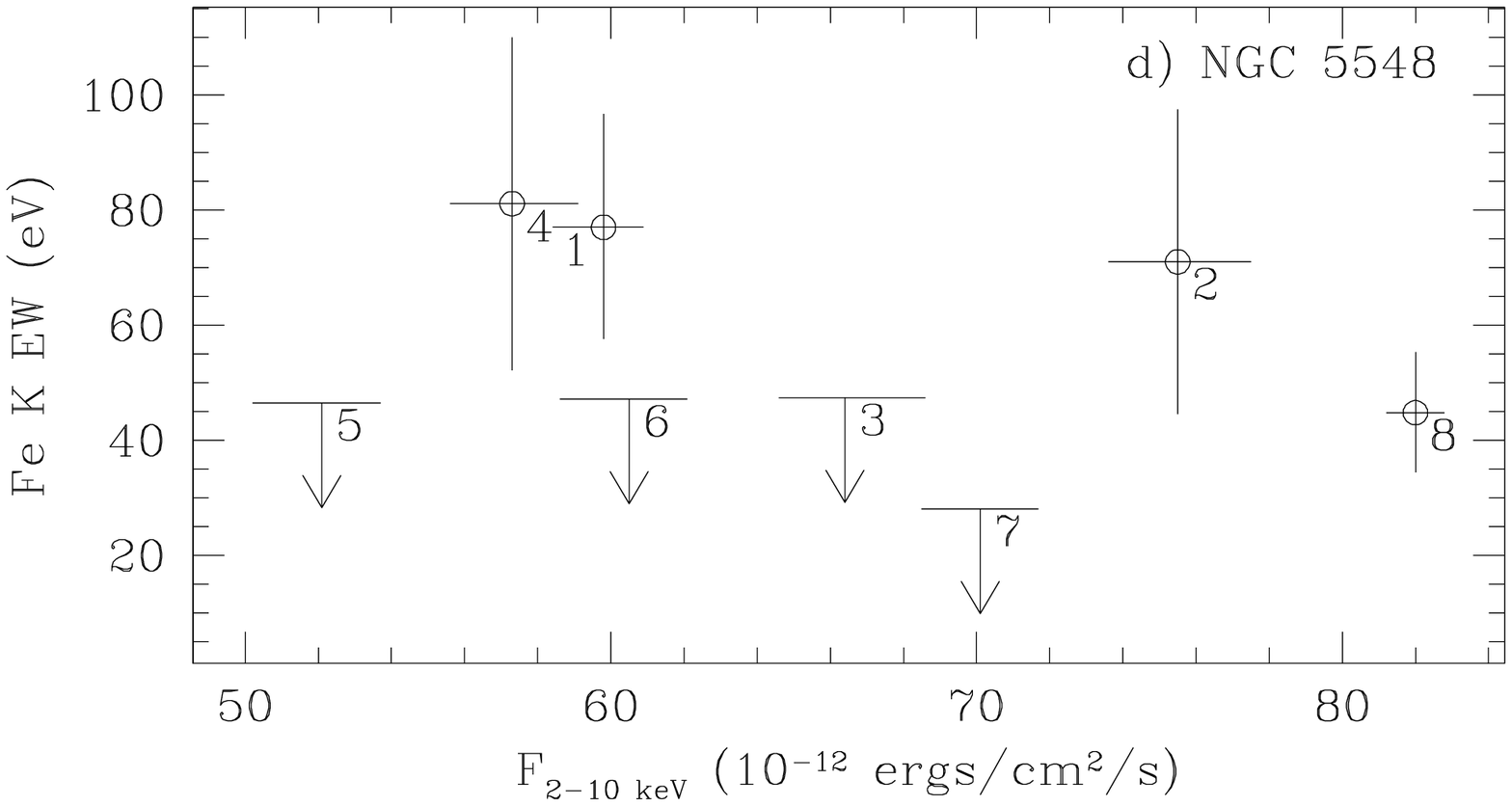} \\
\caption{Line core energy (upper panel) and EW (lower panel) vs.\ 
        2--10~keV flux for \mbox{(a) Fairall 9,} \mbox{(b) NGC 4151,}
        and \mbox{(c) MCG --6-30-15.}
        \mbox{(d) NGC 5548:} Line core EW vs.\ 2--10~keV flux.
        Numeric labels indicate the sequence of observations.
\label{fig:eqw_var}
}
\end{figure}

\addtocounter{figure}{1}

%\clearpage

\begin{figure}
\epsscale{0.7}
\plotone{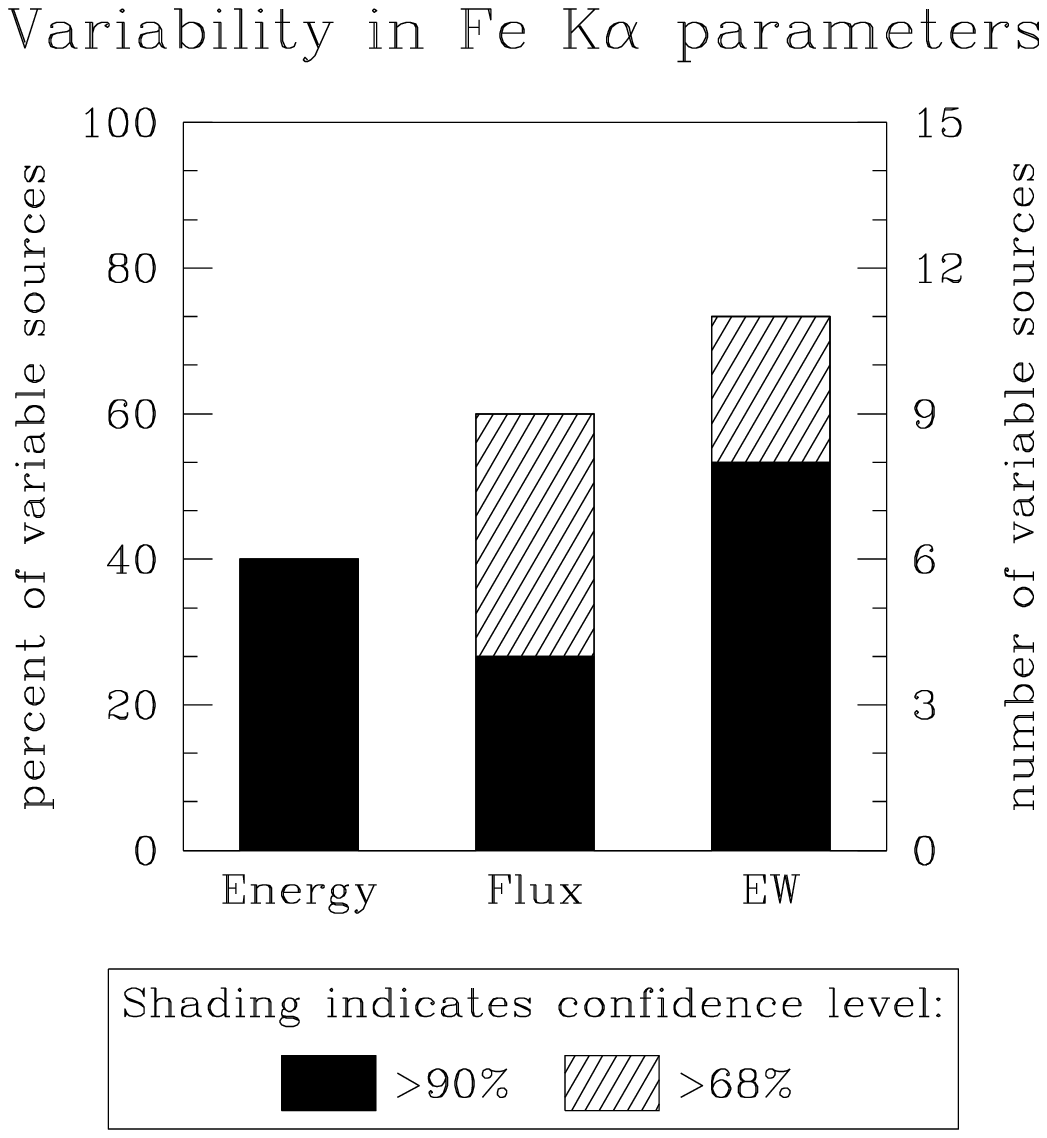}
\caption{
The frequency of Fe~K$\alpha$ variability.
The histogram bars represent the percent of our sample with evidence
for variability in the line core center (energy), flux, and equivalent
width, respectively from left to right.  The filled levels represent
results with $>$$90\%$ confidence, while the diagonally shaded levels
show the $>$$68\%$ results. 
\label{fig:summary}}
\end{figure}

%\clearpage

\begin{figure}
\epsscale{1.0}
\plotone{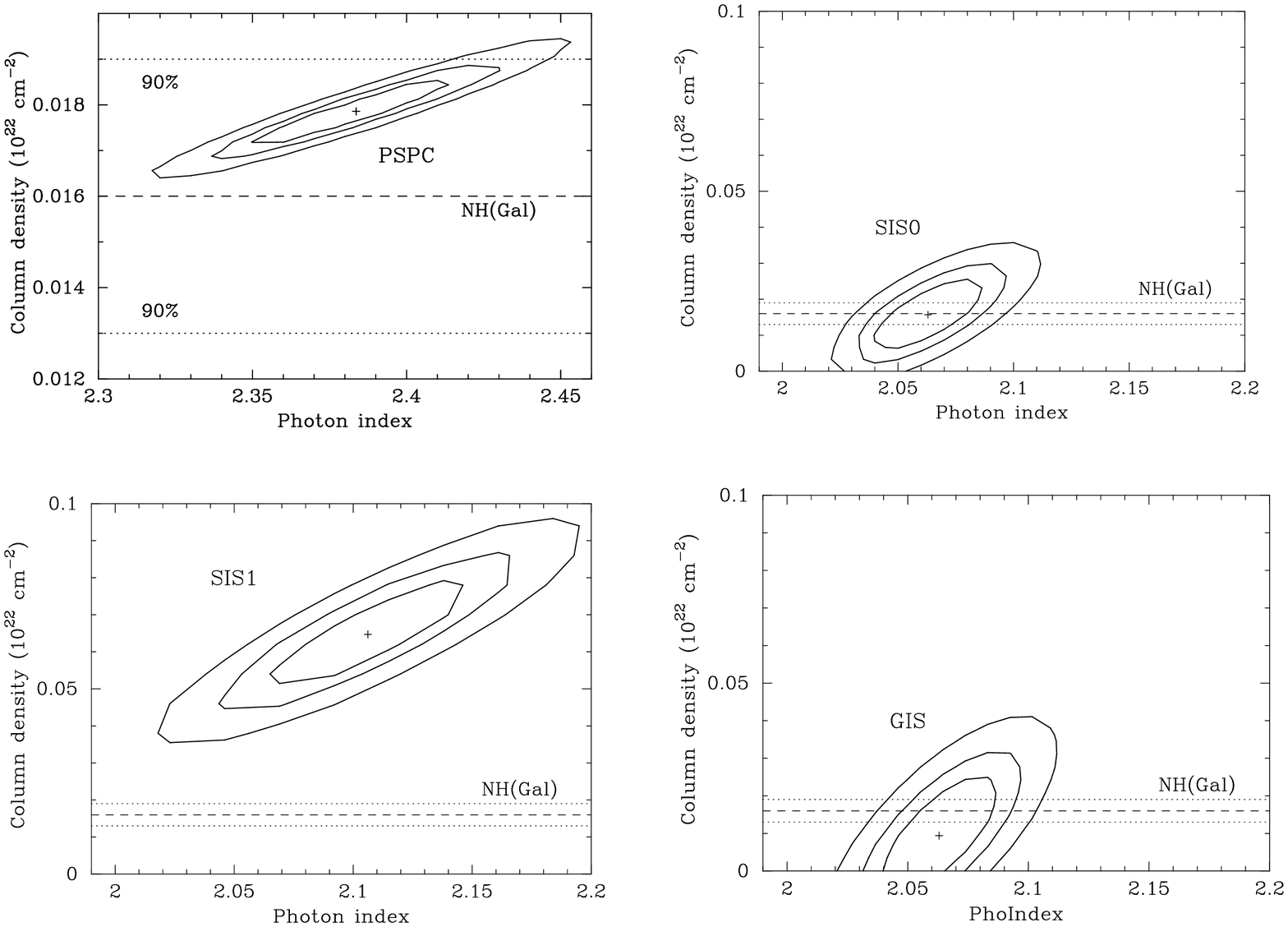}
\caption{$\Gamma$ vs. \NH\ for fits to the Mkn~279 data for (a)~the
{\it Rosat} PSPC, (b)~{\it ASCA} S0, (c)~{\it ASCA} S1 and
(d)~{\it ASCA} GIS (both detectors).  Fits include data from 0.6 to
5~keV.  Contours represent the 68, 90, and $99\%$ confidence limits (for 
two interesting parameters).
\label{fig:contours}
}
\end{figure}

%\clearpage

\begin{figure}
\epsscale{0.8}
\plotone{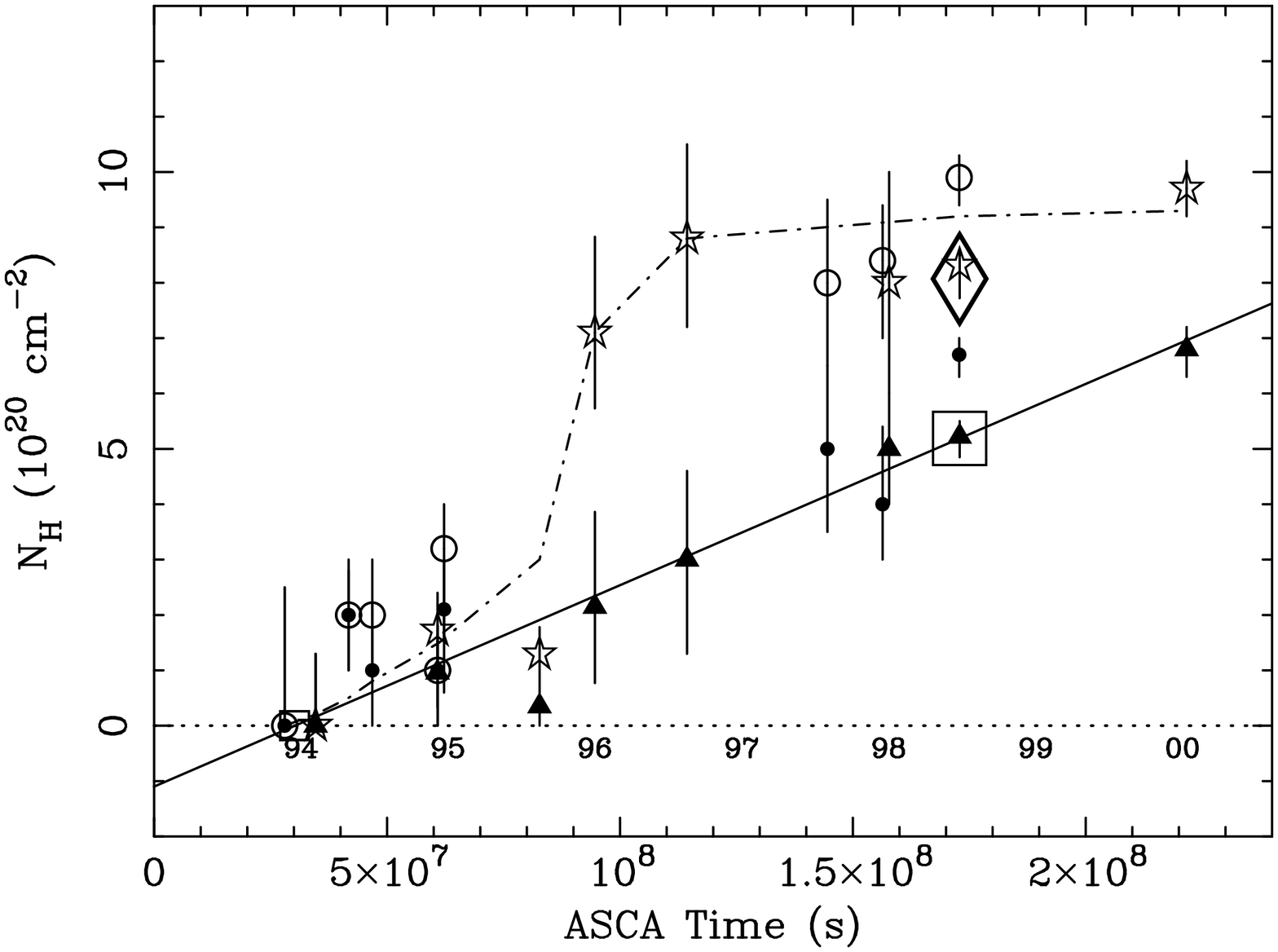}
\caption{Parameterization of ASCA SIS low-energy problem with excess $N_{\rm H}$
(1-CCD mode data) as a function of time.  Solid circles are SIS0
compared with the GIS via model fitting, solid triangles 
are SIS0 compared with the GIS using data ratios,
unfilled circles are SIS1 compared with the GIS
via model fitting, stars are SIS1 compared with the GIS 
using data ratios.  The large square and triangle are 
comparisons of SIS0 with SAX and SIS1 with SAX, respectively. 
The small square is the 1993 Dec 16 calibration point for 3C~273.
The solid line is the best-fitting linear relation for excess $N_{\rm H}$ as a 
function of time for SIS0.  The dot-dashed curve is not a fit, but is 
drawn to guide the eye for the approximate trend for SIS1
with time.
\label{fig:extraNh}
}
\end{figure}

\end{document}